\documentclass[preprint,12pt, authoryear]{elsarticle}

\usepackage{graphicx}
\usepackage{amssymb}
\usepackage{amsthm}
\usepackage{bm}
\usepackage{subfigure}
\usepackage{url}
\usepackage{todonotes}

\usepackage{lineno}

\newcommand{\argmin}{\mathop{\rm arg~min}\limits}

\journal{}

\begin{document}

\begin{frontmatter}


\title{A versatile anomaly detection method for medical images with a flow-based generative model in semi-supervision setting}

%

\author{Hisaichi Shibata\corref{cor1}\fnref{add1}}
\ead{sh-tky@umin.ac.jp}
\author{Shouhei Hanaoka\fnref{add2}}
\author{Yukihiro Nomura\fnref{add1}}
\author{Takahiro~Nakao\fnref{add1}}
\author{Issei Sato\fnref{add6,add2,add5}}
\author{Daisuke~Sato\fnref{add3}}
\author{Naoto Hayashi\fnref{add1}}
\author{Osamu Abe\fnref{add2,add3}}

\cortext[cor1]{Corresponding Author}
\address[add1]{Department of Computational Diagnostic Radiology and Preventive Medicine, The~University of Tokyo Hospital, 7-3-1 Hongo, Bunkyo-ku, Tokyo 113-8655, Japan\fnref{label3}}
\address[add2]{Department of Radiology, The~University of Tokyo Hospital, 7-3-1 Hongo, Bunkyo-ku, Tokyo 113-8655, Japan}
\address[add6]{Department of Computer Science, Graduate School of Information Science and Technology, The~University of Tokyo, 7-3-1 Hongo, Bunkyo-ku, Tokyo 113-0033, Japan}
\address[add5]{Center for Advanced Intelligence Project, RIKEN, 1-4-1 Nihonbashi, Chuo-ku, Tokyo 103-0027, Japan}
\address[add3]{Division of Radiology and Biomedical Engineering, Graduate School of Medicine, The~University of Tokyo, 7-3-1 Hongo, Bunkyo-ku, Tokyo 113-8656, Japan}

\begin{abstract}
Oversight in medical images is a crucial problem, and timely reporting of medical images is desired.
Therefore, an all-purpose anomaly detection method that can detect virtually all types of lesions/diseases in a given image is strongly desired.
However, few commercially available and versatile anomaly detection methods for medical images have been provided so far.
Recently, anomaly detection methods built upon deep learning methods have been rapidly growing in popularity, and these methods seem to provide reasonable solutions to the problem.
However, the workload to label the images necessary for training in deep learning remains heavy.
In this study, we present an anomaly detection method based on two trained flow-based generative models.
With this method, the posterior probability can be computed as a normality metric for any given image.
The training of the generative models requires two sets of images: a set containing only normal images and another set containing both normal and abnormal images without any labels.
In the latter set, each sample does not have to be labeled as normal or abnormal; therefore, any mixture of images (e.g., all cases in a hospital) can be used as the dataset without cumbersome manual labeling.
The method was validated with two types of medical images: chest X-ray radiographs (CXRs) and brain computed tomographies (BCTs).
The areas under the receiver operating characteristic curves for logarithm posterior probabilities of CXRs (0.868 for pneumonia-like opacities) and BCTs (0.904 for infarction) were comparable to those in previous studies with other anomaly detection methods. 
This result showed the versatility of our method.
\end{abstract}


\end{frontmatter}

\section{Introduction}
\label{sec:Introduction}
Radiologists interpret many medical images, and timely reporting for these images is desired \citep{tang2020automated}.
Therefore, an all-purpose anomaly detection method that can detect virtually all types of lesions/diseases in a given image is strongly desired.
However, few commercially available and versatile anomaly detection methods for medical images have been provided so far.
The features that should appear in the desired method are as follows.
(i) It can handle any pathologies anticipated, (ii) it can be trained with datasets that are comparatively easy to collect, e.g., a large number of arbitrary images in hospitals, and with labels that are comparatively easy to input, and (iii) it has versatility for medical images acquired with different modalities. 
In this study, we focus on chest X-ray radiographs (CXRs) and brain computed tomographies (BCTs) as primary targets for anomaly detection.
In this study, anomaly detection is defined as detecting findings different from normal cases.

Recently, with the highest image recognition performance among the machine learning methods, anomaly detection methods based on deep learning (DL) \citep{pang2020deep} have become increasingly popular, including in medical image analysis \citep{litjens2017survey,sahiner2019deep,Kermany2018}.
In the following subsections, a brief review of anomaly detection methods based on DL is presented from three viewpoints: the categories of anomaly detection methods for general images (Sub-sec. \ref{sec:general_images}), the previous anomaly detection strategies for CXRs (Sub-sec. \ref{sec:anomaly_cxrs}), and those for BCTs (Sub-sec. \ref{sec:anomaly_ctbs}).

\subsection{Categories of anomaly detection methods for general images}
\label{sec:general_images}
There are several categories of anomaly detection methods.
In this study, following \citep{NIPS2019_9259}, the categories are defined as follows.
In a supervised setting, a function $\bm{f}_\theta \left( \bm{x}_i \right)$, which forms a mapping from an input vector to an output vector by a deep neural network (DNN), is trained so that the DNN can predict the label of the $i$th input image ($\bm{x}_i \in \mathbb{R}^{H\times W}$ for two-dimensions, $\bm{x}_i \in \mathbb{R}^{D\times H\times W}$ for three dimensions, where $D$, $H$, and $W$ are depth, height, and width of the image respectively) as the output vector.
Therefore, the training of the function is finished when the following $\hat{\theta}$ is found:
\begin{equation}
    \hat{\theta} = \argmin_{\theta} \sum_i L \left[ \bm{f}_{\theta} \left( \bm{x}_i \right), \bm{y}_i \right],
\end{equation}
where $\theta$ represents parameters including weights and biases in the DNN, $\bm{y}_i$ represents labels (e.g., one-hot vector) added to the input images,
and $L$ represents an objective function in the DNN, e.g., mean square, to be minimized.
In a weakly supervised setting, the labels $\bm{y}_i$ cannot be explicitly obtained, e.g., localization results for images are not provided, but classes of the images, e.g., normal or non-normal, are only provided against a DNN for localization purposes.
In a semi-supervised setting, the labels $\bm{y}_i$ can be obtained only for a partial set of input images, e.g., there is a certain image set in which any label is not added.
In an unsupervised setting, no labels are given to all the input images.
Following this definition, the method proposed in this study is classified as a semi-supervised DL method.

\subsection{Anomaly detection methods for CXRs}
\label{sec:anomaly_cxrs}
For CXRs, there are comparatively large-scale datasets open to the public \citep{shih2019augmenting,wang2017chestx}, and several previous studies handled datasets to measure the discriminative performance of a DNN.
Note that the Radiological Society of North America (RSNA) Pneumonia Detection Challenge dataset \citep{shih2019augmenting} is a refined version of the Chest X-ray 8 or 14 dataset \citep{wang2017chestx}.
Because the RSNA Pneumonia Detection Challenge dataset is adopted in this study, this discussion is limited to studies with these two datasets from the viewpoint of discriminative performance comparison. 

For a discriminative model, Rajpurkar et al. \citep{rajpurkar2018deep} established a network called CheXNeXt, and they compared its performance with that of radiologists.
The area under the receiver operating characteristic curve (AUC) obtained with the network was higher than that of radiologists when detecting some pathologies.
Tang et al. \citep{tang2020automated} applied discriminative DNNs to various datasets, including the RSNA Pneumonia Detection Challenge dataset \citep{shih2019augmenting}.
They obtained an AUC of 0.9804 for pneumonia-like lung opacity and an AUC of 0.9492 for other forms of lung opacity by adopting the VGG-19 network.

For a generative model, Tang et al. \citep{Tang2019} adopted a generator built upon an autoencoder in conjunction with a discriminator inspired by generative adversarial networks (GANs) \citep{YI2019101552}.
They achieved a one-class classification by training with only normal images.
They attained an AUC of 0.841 for the Chest X-ray 8 dataset \citep{wang2017chestx}.

\subsection{Anomaly detection methods for BCTs}
\label{sec:anomaly_ctbs}
One of the latest studies with supervised learning for BCTs achieved an AUC of 0.99 for detecting acute intracranial hemorrhage \citep{kuo2019expert}.
This value is an improvement of that obtained by Prevedello et al. \citep{prevedello2017automated}, who first adopted the supervised GoogLeNet, which is composed of two-dimensional convolutional neural networks (CNNs).
At that time, an AUC of 0.91 was obtained for the detection of hemorrhage, mass effect, or hydrocephalus, and an AUC of 0.81 was obtained for the detection of suspected acute infarction.

Titano et al. \citep{titano2018automated} first adopted pure three-dimensional CNNs to classify BCTs in a weakly supervised context.
They obtained an AUC of 0.73 for predicting labels in a certain set.
They annotated BCTs with a semi-supervised natural-language processing (NLP) framework, which contributed to reducing the workload for the labeling.
Another work on supervised DL methods with the aim of reducing the computational cost of handling three-dimensional images was reported by Patel et al. \citep{ISI:000477864400132}, who adopted two-dimensional CNNs in conjunction with bidirectional long short-term memory (LSTM).

As a semi-supervised DL method in which only labels for normal BCTs are required,
Sato et al. \citep{sato2018primitive} proposed a semi-supervised DL method based on a three-dimensional autoencoder.
The autoencoder learns a function $\bm{f}\left(\bm{x}_{i,j} \right)$, which is expressed with multiple CNNs so that $ \sum_{i,j} \left\|\bm{f}\left(\bm{x}_{i,j}\right) - \bm{x}_{i,j} \right\|_2$ is minimized for a training dataset only with normal cases, where $\bm{x}_{i,j}$ is the $j$th patch of the $i$th input image vector, and $\bm{f}$ represents an autoencoder-decoder.
If a test image ($\bm{x}_{i,j}$) with an abnormality is passed through the trained autoencoder, the metric values $ \left\|\bm{f}\left(\bm{x}_{i,j}\right) - \bm{x}_{i,j} \right\|_2$, are expected to be larger than those with test images without an abnormality.
On the basis of this fact, they introduced a heuristic function, $J_i = \max\limits_{j} \|\bm{f}\left(\bm{x}_{i,j}\right) - \bm{x}_{i,j}\|_2$, as a normality metric for the $i$th test image and obtained an AUC of 0.87.
This method can potentially handle BCTs with unknown pathologies that did not exist when the autoencoder was trained.
Note that the method proposed in this study has a similar capability.

\subsection{The objectives of this study}
The objective of this study is to develop an anomaly detection method built upon deep learning with semi-supervision.
The method is realized by combining two flow-based generative models, with which the computation of a logarithm posterior probability is possible.
We adopt Glow \citep{Kingma2018} to estimate a probability density function for any given medical images.
Glow is one of the flow-based generative models with which realistic fictional image generation was first achieved.
To show the versatility of the proposed method for different types of medical images, we validate the method with CXRs and BCTs using different models.
The dataset for CXRs is taken from the RSNA Pneumonia Detection Challenge dataset \citep{shih2019augmenting}, and the dataset for BCTs is collected from the emergency department of our institution.

The rest of this paper is organized as follows.
In Sec.~\ref{sec:methods}, the method used to compute a normality metric for anomaly detection in CXRs or BCTs is introduced.
The setting and the dataset used to execute numerical experiments are reported in Sec.~\ref{sec:exp_cxr} for CXRs and in Sec.~\ref{sec:exp_ctb} for BCTs.
In Sec.~\ref{sec:results}, the experimental results for both CXRs and BCTs are given.
In Sec.~\ref{sec:discussion}, a discussion based on the experimental results is presented.
In Sec.~\ref{sec:conclusions}, the study is concluded.

\section{Methods}
\label{sec:methods}
To judge the normality of a medical image given ($\bm{x}_i$), the posterior probability that the image is normal, $p(\mathrm{normal}|\bm{x}_i)$, is evaluated using two learned flow-based generative models.
\begin{figure}
    \centering
    \includegraphics[keepaspectratio,scale=0.8]{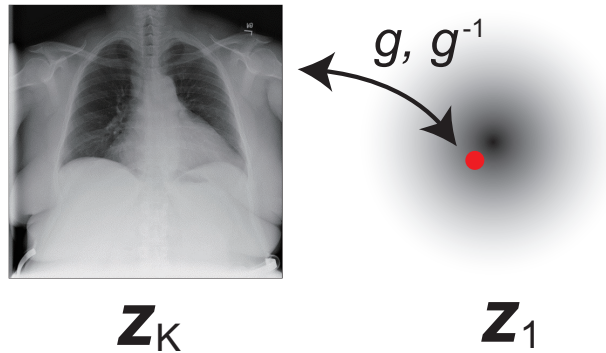}
    \caption{Overview of the flow-based DNN, where $g= f_1 \circ f_2 \circ \cdots \circ f_{K-1}$ and the red point indicates the destination of the transformation onto the high-dimensional multivariate Gaussian distribution.}
    \label{fig:flow}
\end{figure}
The concept of flow-based DNNs is shown in Fig.~\ref{fig:flow}.
In flow-based DNNs, space variables corresponding to input images ($\bm{x}_i$) are recursively projected onto a latent space variable ($\bm{z}_i \in \mathbb{R}^{H\times W}$ for 2-D, $\bm{z}_i \in \mathbb{R}^{D\times H\times W}$ for 3-D; hereafter, we show the 2-D case) using invertible transformations $\bm{z}_{i, k+1} = \bm{f}_k \left( \bm{z}_{i,k} \right)$ for $k=1,\cdots, \left(K-1\right)$, where $K$ is a constant, $\bm{z}_{i,K} = \bm{x}_i$, and $\bm{z}_{i,1} \in \mathbb{R}^{H\times W}$ is targeted as a point typically in a high-dimensional multivariate Gaussian distribution.
The transformation of logarithm probability density functions between that of the targeted distribution and that of the input image, e.g., a CXR or a BCT, is given by
\begin{eqnarray}
    \log p \left( \bm{x}_i \right)\triangleq \log{p\left( \bm{z}_{i,K} \right)} =   \log{p\left( \bm{z}_{i,1} \right)} + \sum_{k=1}^{K-1} \log{\mathrm{det}\left|\left( \frac{\mathrm{d}\bm{f}_{k}}{\mathrm{d}\bm{f}_{k-1}}\right)\right|},
\end{eqnarray}
where $\bm{f}_0 \triangleq  \bm{x}_{i}$ and $\bm{f}_{K-1} \triangleq \bm{z}_{i,1}$.
Note that, with the above transformation formula, the likelihood of any given image for a model can be computed explicitly, unlike with other generative models, e.g., GANs and variational autoencoders.
For details of the flow-based generative model, see \citep{Kingma2018}.

Different from the discriminative model, with which the posterior probability is directly available, the probability is computed with Bayes' theorem in the flow-based generative model as follows:
\begin{eqnarray}
    \log p\left(\mathrm{normal} |\bm{x}_i\right) &=& \log p\left(\bm{x}_i|\mathrm{normal} \right) +\log p\left(\mathrm{normal} \right) - \log p\left(\bm{x}_i \right),
    \label{eqn:beyes}
\end{eqnarray}
where $p\left(\bm{x}_i | \mathrm{normal}\right)$ is the probability distribution function (PDF) for the population of normal samples, $p\left( \bm{x}_i \right)$ is the PDF for the population of all samples, and $p\left(\mathrm{normal}\right)$ is the appearance probability of normal samples in the population, which takes a constant value for a fixed population and does not affect the receiver operating characteristic (ROC) curves drawn in the testing.
Therefore, the constant is omitted in the following to enhance readability.
Note that a similar formulation was introduced in \citep{Ren2019,serra2019input}. 

To compute the logarithm posterior probability, the preparation of three sets is a prerequisite: a medical image set, $\mathcal{S}_\mathrm{normal}^{\mathrm{train}}$, which only contains normal medical images for training,
and two other medical image sets, $\mathcal{S}^\mathrm{train}_\mathrm{mixture}$ and $\mathcal{S}^\mathrm{test}_\mathrm{unknown}$, which contain both normal and non-normal medical images, for training and testing, respectively.
For conciseness, in the rest of this section, we hypothesize that the probability distributions of the second training set $\mathcal{S}^\mathrm{train}_\mathrm{mixture}$ and the test set $\mathcal{S}^\mathrm{test}_\mathrm{unknown}$ are derived from the same population, which represents the real occurrence probabilities of all possible abnormalities (e.g., diseases).

\begin{figure}
    \centering
    \includegraphics[keepaspectratio,scale=0.38]{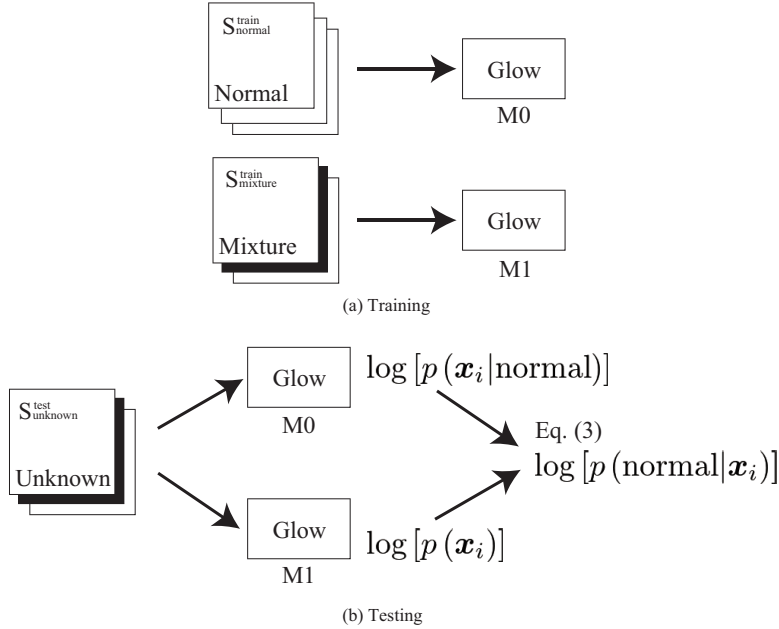}
    \caption{Flowcharts to compute the logarithm posterior probability ($\log p(\mathrm{normal}|\bm{x}_i)$) with the flow-based generative model.}
    \label{fig:procedure}
\end{figure}

Flowcharts to compute the logarithm posterior probability ($\log p(\mathrm{normal}|\bm{x}_i)$) for the training and the testing are respectively shown in Figs.~\ref{fig:procedure}(a) and Fig.~\ref{fig:procedure}(b).
The procedures for computing the logarithm posterior probability ($\log p(\mathrm{normal}|\bm{x}_i)$) are listed as follows:
\begin{enumerate}
	\item Prepare input data:
	\begin{enumerate}
		\item Gather a set of normal medical images, $\mathcal{S}_\mathrm{normal}^\mathrm{train}$
		\item Gather a set of a mixture of normal and non-normal medical images, $\mathcal{S}^\mathrm{train}_\mathrm{mixture}$ and $\mathcal{S}^\mathrm{test}_\mathrm{unknown}$
	\end{enumerate}
	\item Train a flow-based generative model ($\mathcal{M}_0$) with $\mathcal{S}_\mathrm{normal}^\mathrm{train}$ to maximize $\sum_i \log{p\left(\bm{x}_i|\mathrm{normal}\right)}, \forall \bm{x}_i \in \mathcal{S}_\mathrm{normal}^\mathrm{train}$
	\item Train another flow-based generative model ($\mathcal{M}_1$) with $\mathcal{S}^\mathrm{train}_\mathrm{mixture}$ to maximize $\sum_i \log{p\left(\bm{x}_i\right)}, \forall \bm{x}_i \in \mathcal{S}^\mathrm{train}_\mathrm{mixture}$
	\item Infer with $\mathcal{M}_0$ using medical images for testing ($\forall \bm{x}_i \in \mathcal{S}^\mathrm{test}_\mathrm{unknown} $) to obtain $\log p\left(\bm{x}_i |\mathrm{normal}\right)$
    \item Infer with $\mathcal{M}_1$ using medical images for testing ($\forall \bm{x}_i \in \mathcal{S}^\mathrm{test}_\mathrm{unknown}$) to obtain $\log p\left(\bm{x}_i\right)$
    \item Compute $\log p\left(\mathrm{normal}|\bm{x}_i \right)$ for all the test medical images using Eq.~\ref{eqn:beyes}.
\end{enumerate}

To compute the logarithm posterior probability ($\log p \left( \mathrm{normal}|\bm{x}_i\right)$) concretely, the implementation of an estimator for the two kinds of logarithm likelihoods ($p(\bm{x}_i)$ and $\log p(\bm{x}_i|\mathrm{normal})$) is required.
In this study, the flow-based generative model Glow \citep{Kingma2018} is adopted as the estimator.
In addition, with Glow, it is expected that realistic fictional images are generated using random sampling with a trained model.

Furthermore, note that the logarithm likelihood ($\log p\left(\bm{x}_i | \mathrm{normal}\right)$) is simultaneously obtained during the computation of the logarithm posterior probability (see Fig.~\ref{fig:procedure}).
Because the logarithm likelihood is expected to reflect the normality of an input image, we experimented with another anomaly detection method using the logarithm likelihood.
See Appendix A for the details of results for anomaly detection with the logarithm likelihood.

\section{Experiments for CXRs}
\label{sec:exp_cxr}
\subsection{Experimental data}
The CXRs used in this study were originally taken from the RSNA Pneumonia Detection Challenge dataset \citep{shih2019augmenting}.
This dataset comprises 30,000 frontal-view CXRs, with each image labeled as ``Normal,'' ``No Opacity/Not Normal,'' or ``Opacity'' by one to three board-certified radiologists.
The Opacity group consists of images with suspicious opacities suggesting pneumonia, and the No Opacity/Not Normal group consists of images with abnormalities other than pneumonia.

\begin{table}[htb]
  \begin{center}
  \caption{Composition of the three sets. The numbers in brackets represent the numbers of overlapping normal CXRs in the two sets.}
  \begin{tabular}{lcc} \hline
    Set & Num. normal CXRs & Num. non-normal CXRs \\ \hline 
    $\mathcal{S}_\mathrm{normal}^\mathrm{train}$ & 7,808 (6,529)  & 0 \\
    $\mathcal{S}^\mathrm{train}_\mathrm{mixture}$ & 6,553 (6,529) & 6,631 \\
    $\mathcal{S}^\mathrm{test}_\mathrm{unknown}$ & 1,358 & 13,863 \\ \hline
    \end{tabular}
    \end{center}
    \label{tbl:set_cxr}  
\end{table}
The composition of the CXR sets used in the present experiments is shown in Table~1.
In this study, the labels with the three categories are utilized only for validating inference results, and the training and the testing are executed without explicitly providing any labels. 

\subsection{Experimental setup}
The original Glow code \citep{Kingma2018,glow_url} is adopted in this study with slight modifications:
the variance and the mean of the latent variables are set to 1 and 0, respectively, and the scale function in the affine coupling layer is modified to the scale $s\left( h_2 - 0.1 \right) + 0.6$ from the scale $s\left( h_2 + 2.0\right)$, where $s$ is the sigmoid function and $h_2$ is the input from the previous split layer.
The hyperparameters adjusted for this study are enumerated in Table~2.

\begin{table}[htb]
  \begin{center}
    \caption{Hyperparameters used to execute Glow code}
    \begin{tabular}{lc} \hline
      Coupling layer & Affine \\
      Flow permutation & 1$\times$1 convolution \\
      Mini-batch size & 128 \\
      (Recurrent) levels & 7 \\
      Depth per level & 32 \\
      Image size (in pixel) & H$512 \times $ W$512 \times$ C1 \\
      Total epochs & 200 \\ 
      Learning rate in steady state& $10^{-3}$ \\\hline 
    \end{tabular}
    \end{center}
    \label{tab:hps}
\end{table}

We utilize Tensorflow 1.12.0 for the back-end of the DNNs.
The versions of CUDA and cuDNN used are 9.0.176 and 7.4, respectively. 
All processes for the CXRs are carried out in one computing node of the Reedbush-L supercomputer system in the Information Technology Center, The University of Tokyo.
The system consists of 64 computing nodes equipped with two Intel Xeon E5-2695v4 processors, 256 GB memory, and four GPUs (NVIDIA Tesla P100 SXM2 with 16 GB memory).
With the above settings, the logarithm posterior probability is computed for given test images, and ROC curves are plotted on the basis of the probability after training the models.

\section{Experiments for BCTs}
\label{sec:exp_ctb}
\subsection{Experimental data}
This study was approved by the ethical review board of our institution.
The BCT images used in this study were collected from the emergency department of our institution.
We first selected 3,845 subjects who received a non-contrast head CT scan from April 2015 to July 2018.
The initial inclusion criteria were as follows: 1) the first-time scan at the emergency department of our institution during the period, 2) aged seven years or older, 3) no postoperative or post-radiation on the brain or skull, and 4) existing axial series covering the whole brain with the reconstruction slice thickness and interval of 4 mm.
One board-certified radiologist (D.S., seven years of experience in BCT image interpretation) reviewed the head BCT images and radiology reports.
He excluded 597 images, including artifacts or inappropriate images.
Of the 3,248 images, we utilized 283 images, which include hemorrhages or acute infarctions (hereafter, non-normal) and 1,357 images with no findings (hereafter, normal).
Of the 1,357 normal images, we extracted 533 images for anomaly detection, and 1,355 images for fictional image generation, in ascending order of scan date.
These images were scanned at our institution with two CT scanners (Aquilion and Aquilion PRIME, Canon Medical Systems Corporation, Otawara, Japan).
The acquisition parameters were as follows: number of detector rows, 64 for Aquilion and 160 for Aquilion PRIME; tube voltage, 120 kVp; tube current, 115--400 mA; soft-tissue reconstruction kernel, FC64; reconstruction slice thickness and interval, 4 mm; field of view, 220 mm; matrix size, 512$\times$512 pixels; pixel spacing, 0.43 mm.

To handle the BCTs with limited computational resources while maintaining information of the brain structure, the 16-bit signed integer images ($I_\mathrm{src}$, CT number in HU units) acquired with these two instruments were converted into 7-bit unsigned integer images ($I_\mathrm{dst}$, similar to brain window) with the following formula:
\begin{eqnarray}
    I_\mathrm{dst} = \mathrm{clip}\left(I_\mathrm{src} + 14, 0, 127\right),
\end{eqnarray}
where the operator $\mathrm{clip}(x, a, b)$ restricts the value range of an array $x$ from $a$ to $b$. 

\begin{table}[htb]
  \begin{center}
  \caption{Composition of the three sets. There are no overlapping images for BCTs.}
  \begin{tabular}{lcc} \hline
    Set & Num. normal & Num. non-normal \\ \hline 
    $\mathcal{S}_\mathrm{normal}^\mathrm{train}$ & 250 & 0 \\
    $\mathcal{S}^\mathrm{train}_\mathrm{mixture}$ & 250 & 250 \\
    $\mathcal{S}^\mathrm{test}_\mathrm{unknown}$ & 33 & 33 \\ \hline
    \end{tabular}
    \end{center}
    \label{tbl:set_ctb}  
\end{table}
The composition of the BCT sets used in the present experiments is shown in Table~3.

Furthermore, to confirm that the flow-based generative model can generate a fictional image using random sampling, 1,355 out of the 1,357 normal images are prepared for another training.
These images are augmented seven times by rotating them $\pm 2^\circ$ in each direction.  

\subsection{Experimental setup}
For the BCTs, the original Glow code \citep{glow_url} is extended to 3-D Glow so that three-dimensional images can be handled; see Appendix B for details of the modifications.
The hyperparameters in the applied code are enumerated in Table~4.
\begin{table}[htb]
  \begin{center}
    \caption{Hyperparameters used to execute 3-D Glow code.}
    \begin{tabular}{lc} \hline
      Coupling layer & Affine \\
      Flow permutation & 1$\times$1$\times$1 convolution \\
      Mini-batch size & 1 \\
      (Recurrent) levels & 4 \\
      Depth of network & 32 \\
      Width of hidden layer & 512 \\      
      Image size (in pixels) & D$32 \times$ H$128 \times $W$128 \times$C1\\
      & (for anomaly detection) \\
      &  D$32 \times$ H$64 \times $W$64 \times$C1 \\
      & (for fictional image generation)\\
      Total epochs & 30 (Train 1), 20 (Train 2) \\ 
      Learning rate in steady state & $10^{-4}$ \\\hline
    \end{tabular}
  \end{center}
    \label{tab:hps2}  
\end{table}
The mini-batch size is set to 1 to avoid out-of-memory exceptions.
We utilize Tensorflow 1.14.0 for the back-end of the DNNs.
The versions of CUDA and cuDNN used are 10.0.130 and 7.4, respectively. 
All processes for the BCTs are carried out on a workstation, consisting of two Intel Xeon Gold 6230 processors, 384 GB memory, and five GPUs (NVIDIA Quadro RTX 8000 with 48 GB memory).
For anomaly detection, we utilize only two out of the five GPUs.
With the above settings, the logarithm posterior probability is computed for given test images, and ROC curves are plotted on the basis of the probability after training the models, similarly to the experiment for CXRs.

\section{Results}
\label{sec:results}

\begin{figure}
    \centering
    \includegraphics[keepaspectratio,scale=0.35]{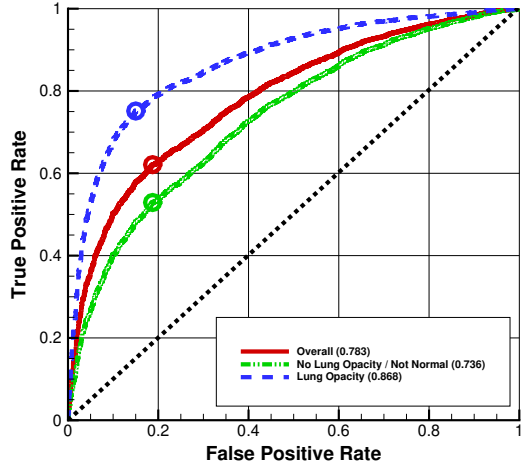}
    \caption{ROC curves obtained with the logarithm posterior probability for CXRs. AUCs for the labels are indicated in brackets, and the circles represent cutoff points obtained from Youden's index.}
    \label{fig:lhr}
\end{figure}
\begin{figure}
    \centering
    \includegraphics[keepaspectratio,scale=0.35]{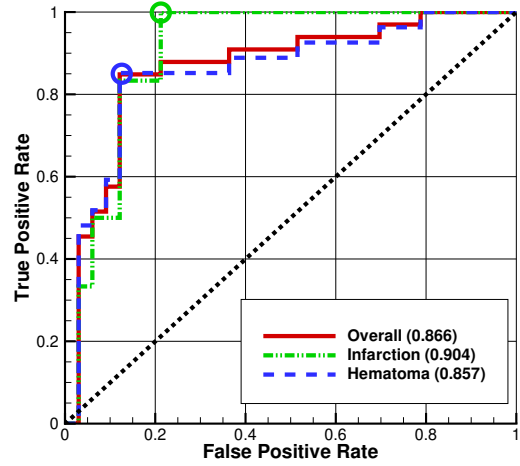}
    \caption{A ROC curve obtained with the logarithm posterior probability for BCTs. AUCs for the labels are indicated in brackets, and the circles represent cutoff points obtained from Youden's index.}
    \label{fig:lhr_ctb}
\end{figure}

ROC curves obtained with the logarithm posterior probability ($\log p\left(\mathrm{normal}|\bm{x}_i \right)$) with different target pathologies are shown in Fig.~\ref{fig:lhr} for CXRs and in Fig.~\ref{fig:lhr_ctb} for BCTs.
The optimal cutoff points for the ROC curves were determined from Youden’s index \citep{youden1950index}.
The AUCs for overall pathologies are approximately 0.783 for CXRs and 0.866 for BCTs.

\begin{figure}
    \centering
    \includegraphics[width=7cm]{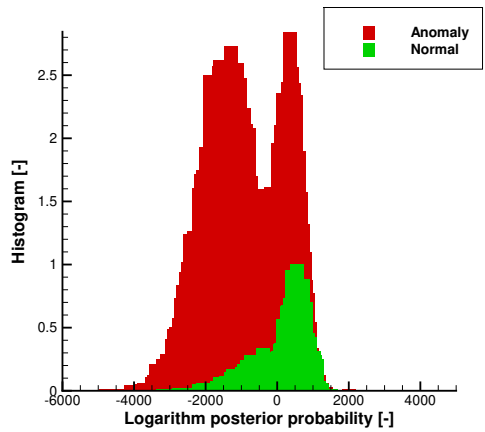}
    \caption{Histogram of the logarithm posterior probability for CXRs.}
    \label{fig:cxr_hist_prob}
\end{figure}

\begin{figure}
    \centering
    \includegraphics[width=7cm]{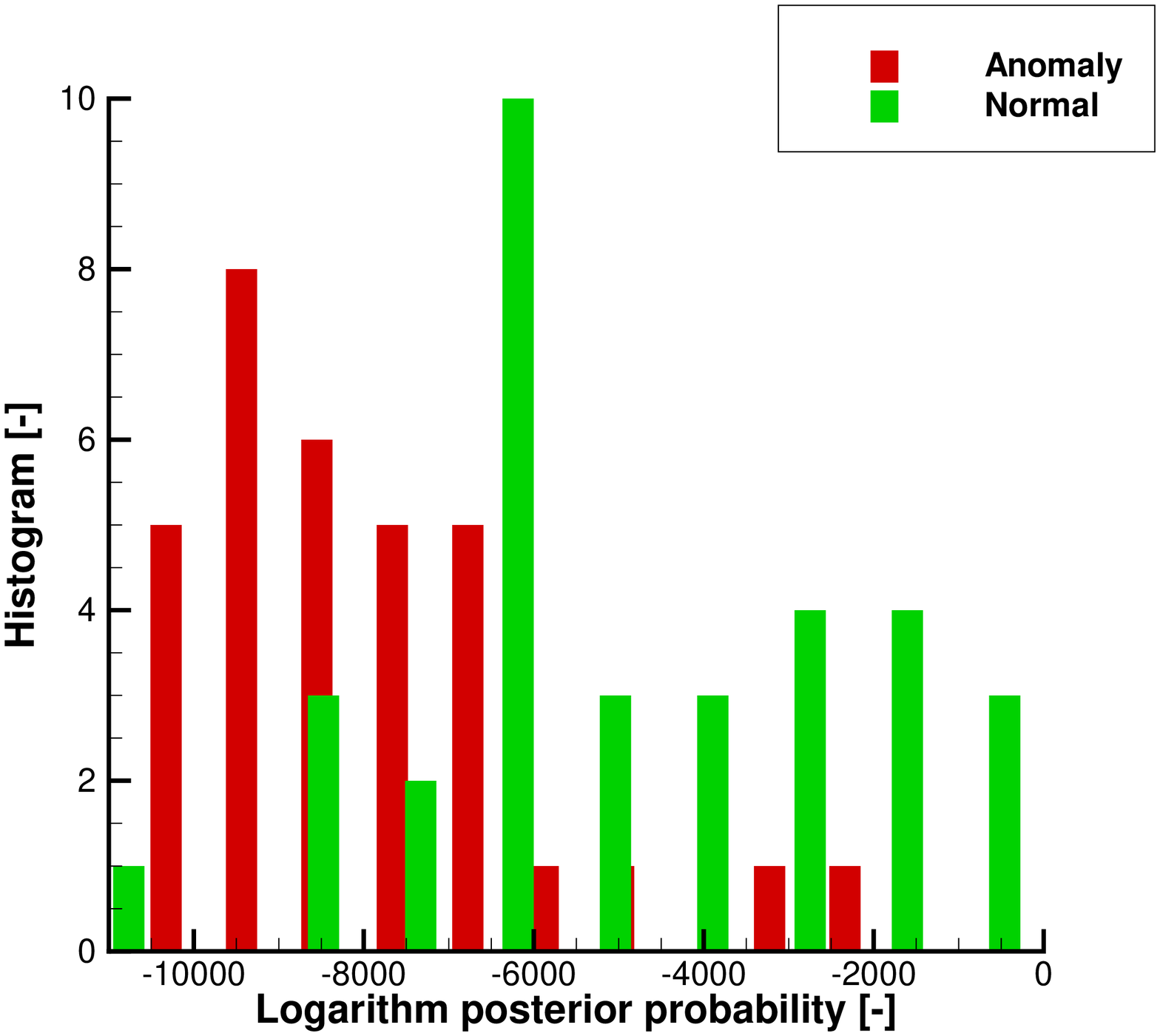}
    \caption{Histogram of the logarithm posterior probability for BCTs.}
    \label{fig:bct_hist_prob}
\end{figure}

Histograms of the logarithm posterior probability are shown in Fig.~\ref{fig:cxr_hist_prob} for CXRs and in Fig.~\ref{fig:bct_hist_prob} for BCTs.



\begin{figure}
    \centering
    \includegraphics[width=14.0cm]{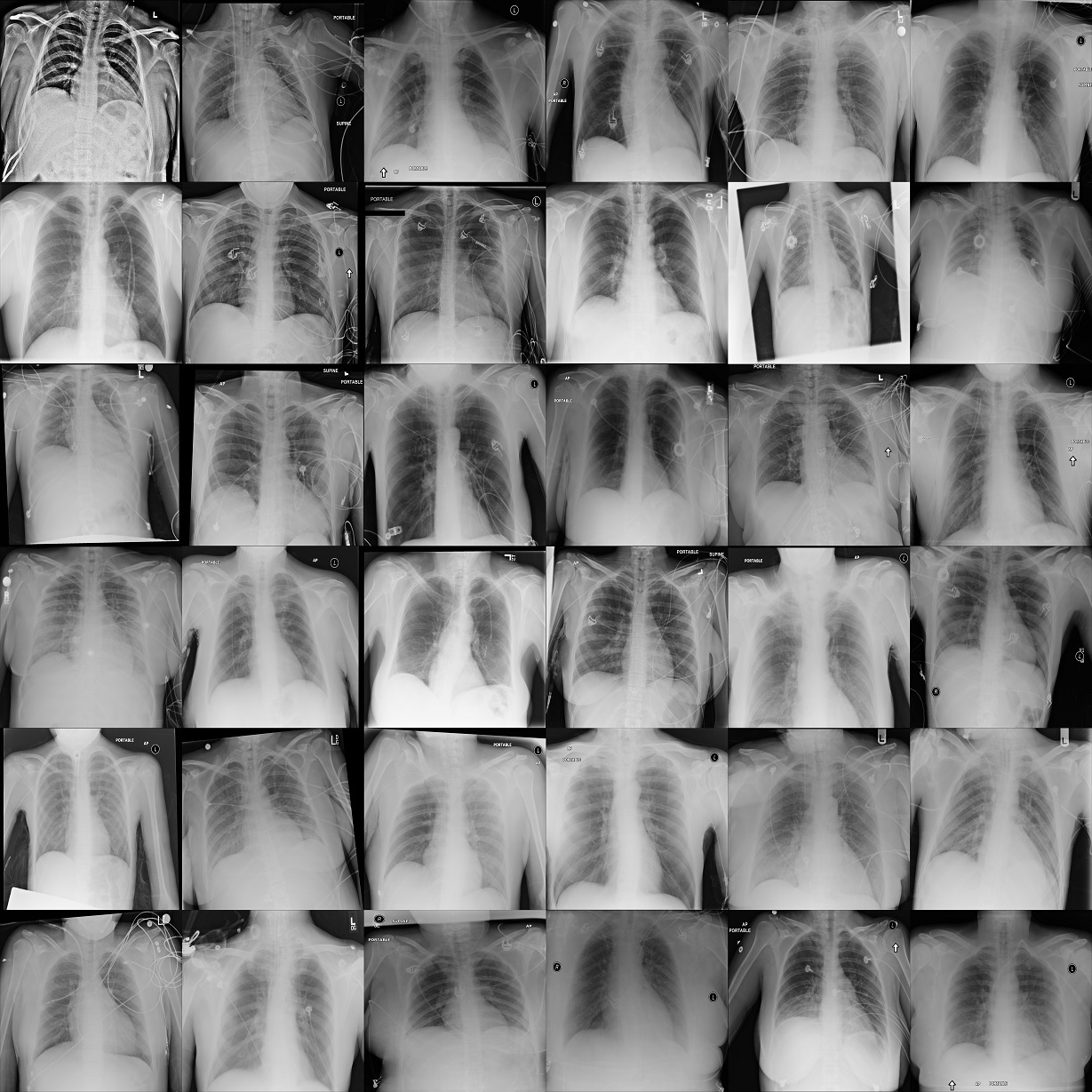}
    \caption{Top 36 CXRs most likely to have an abnormality according to the logarithm posterior probability.}
    \label{fig:multi3of4}
\end{figure}

\begin{figure}
    \centering
    \includegraphics[width=14.0cm]{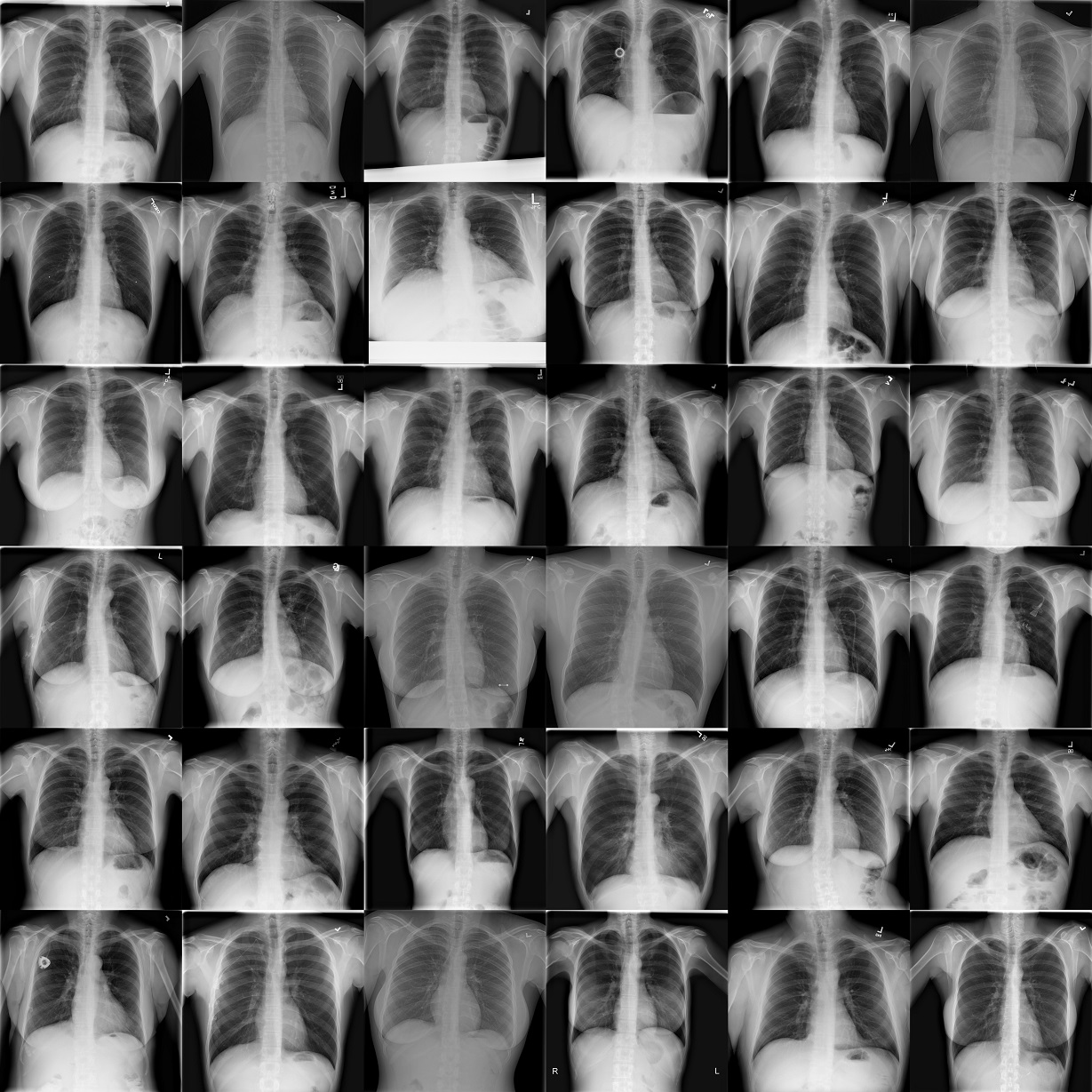}
    \caption{Top 36 CXRs least likely to have an abnormality according to the logarithm posterior probability.}
    \label{fig:multi4of4}
\end{figure}

Figures~\ref{fig:multi3of4} and \ref{fig:multi4of4} show the top 36 CXRs most and least likely to have an abnormality from all the test CXRs, respectively.

\begin{figure}
	\centering
	\subfigure[]{\includegraphics[width=4cm]{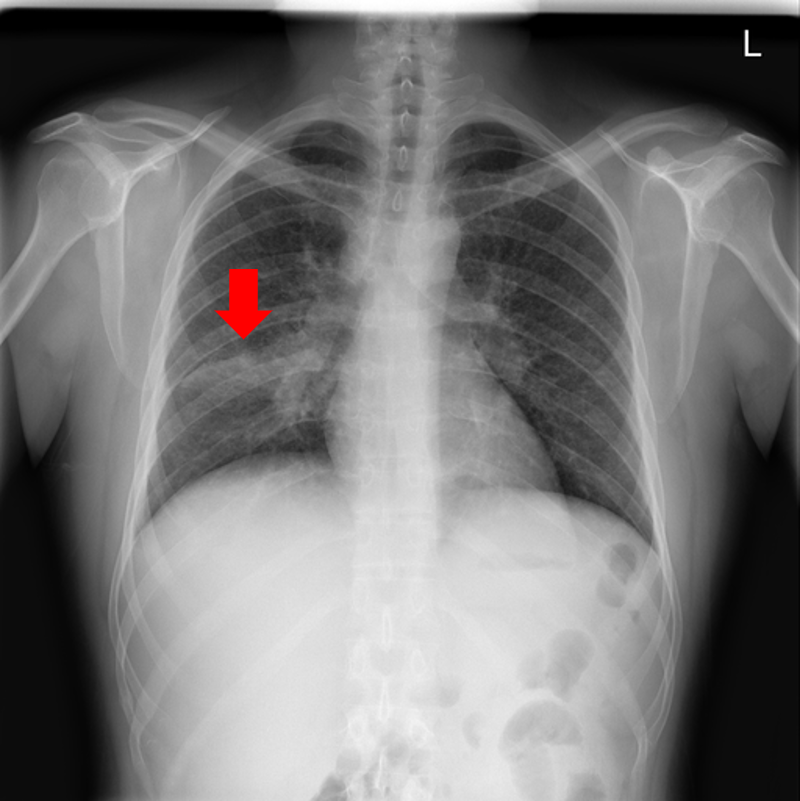}}
	\subfigure[]{\includegraphics[width=4cm]{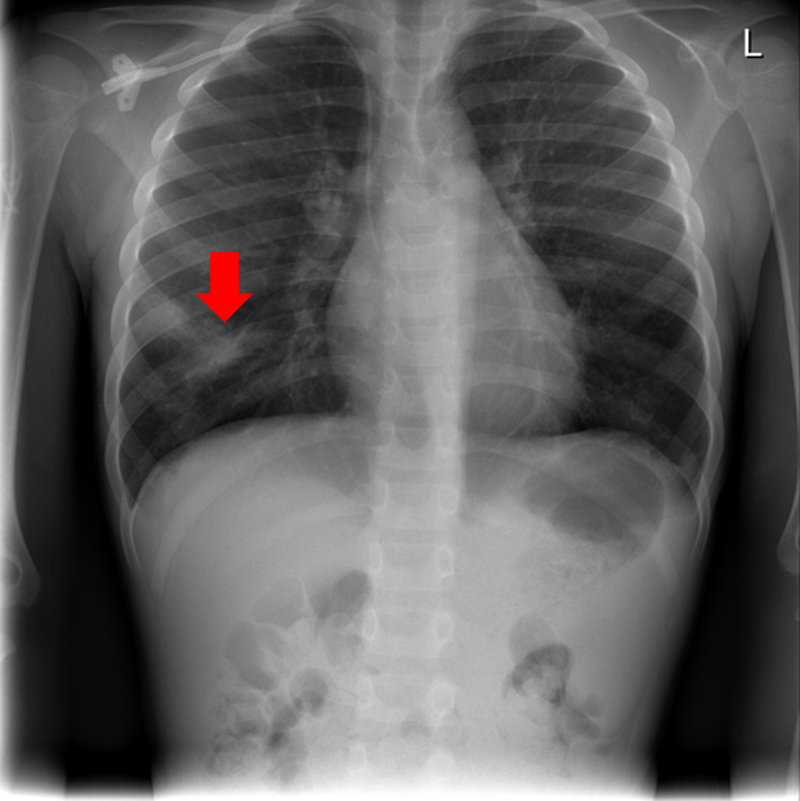}}
	\subfigure[]{\includegraphics[width=4cm]{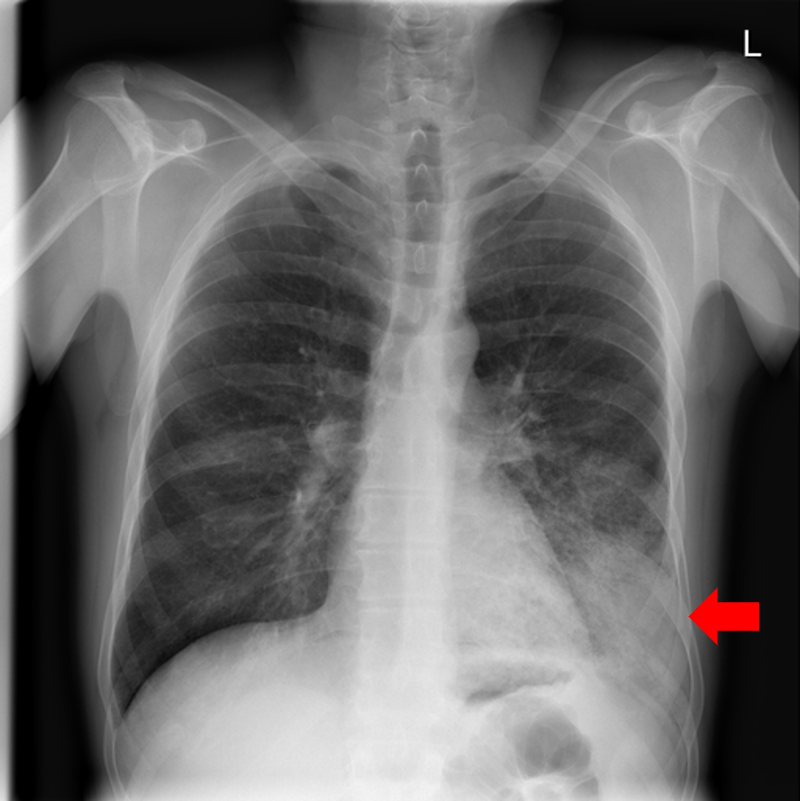}} \\
	\subfigure[]{\includegraphics[width=4cm]{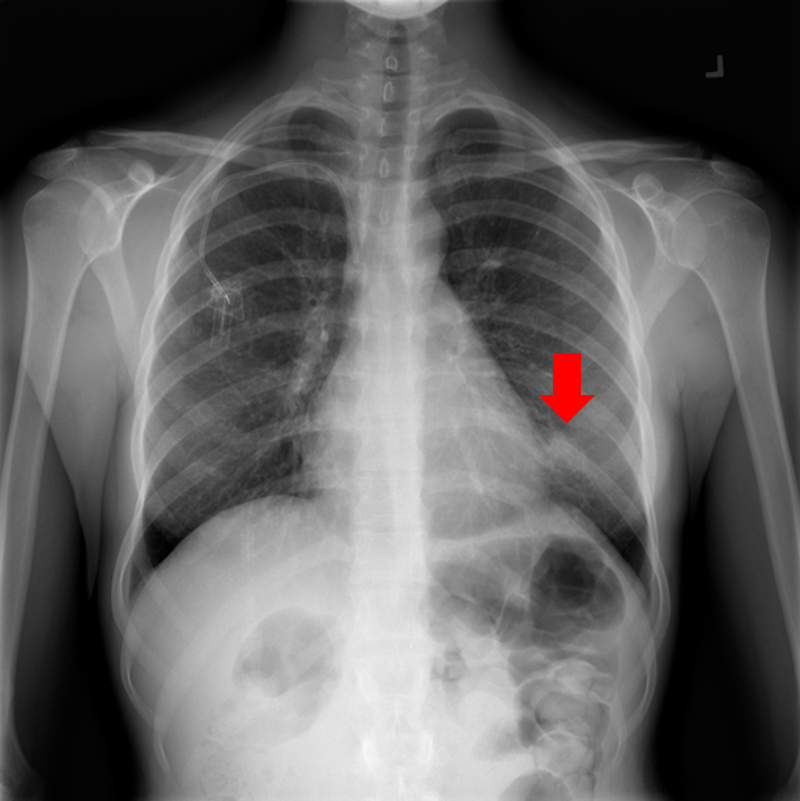}}
	\subfigure[]{\includegraphics[width=4cm]{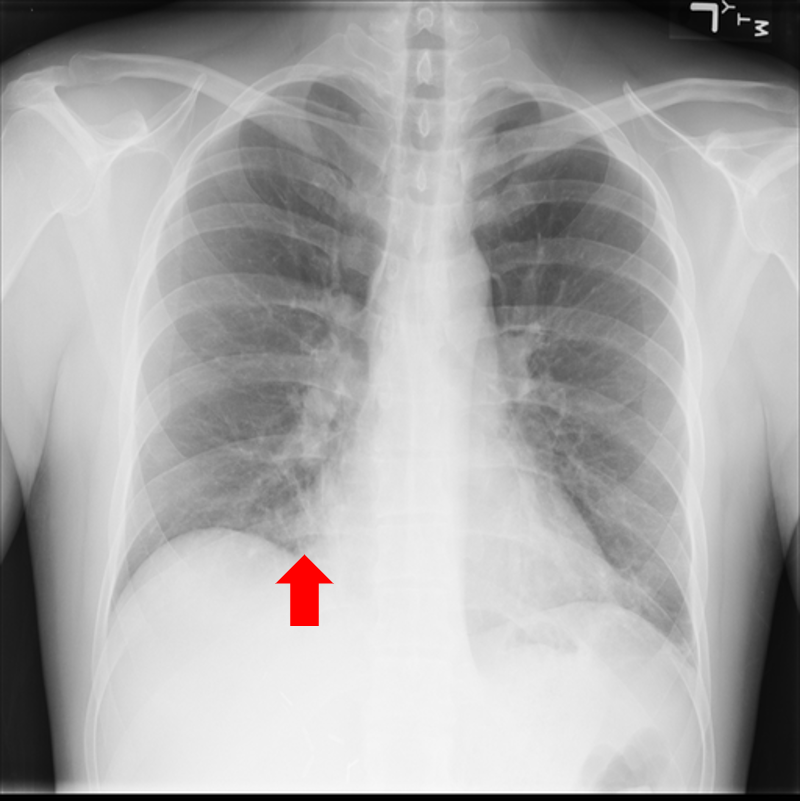}}
	\subfigure[]{\includegraphics[width=4cm]{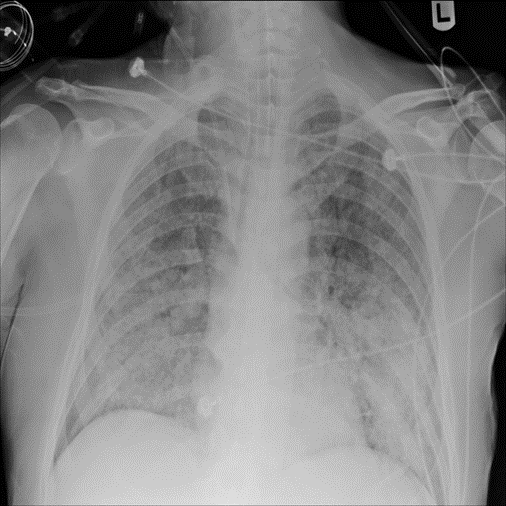}}
	\vspace{-0.2cm}
	\caption{Typical CXRs with pneumonia-like lung opacity from among the top 1,000 CXRs most likely to have an abnormality according to the logarithm posterior probability out of the 13,863 abnormal CXRs. The lesion is indicated with a red arrow if applicable.}
	\label{fig:opacity}
\end{figure}
Figure~\ref{fig:opacity} shows CXRs suggested to have pneumonia-like lung opacity with the logarithm posterior probability.
These CXRs are the top 1,000 CXRs most likely to have an abnormality out of the 13,863 abnormal CXRs.

\begin{figure}
	\centering
	\subfigure[Pleural effusion]{\includegraphics[width=4cm]{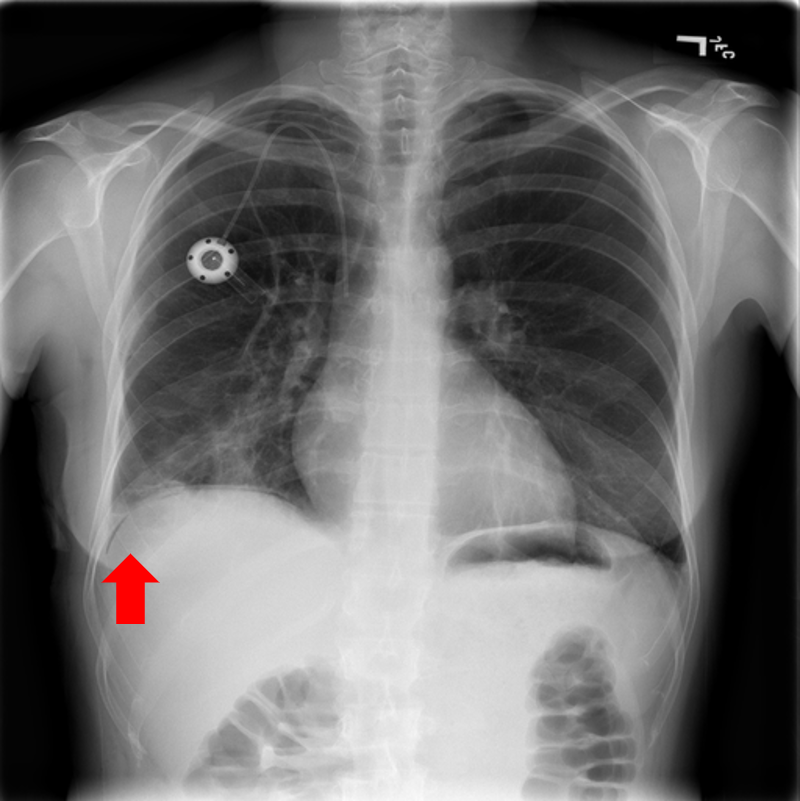}}
	\subfigure[Emphysema]{\includegraphics[width=4cm]{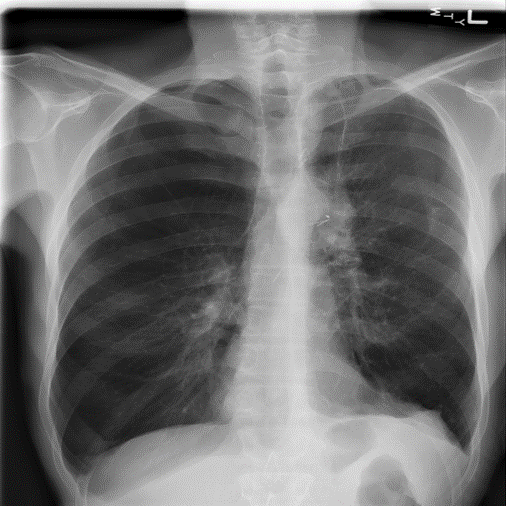}}
	\subfigure[Mediastinal mass]{\includegraphics[width=4cm]{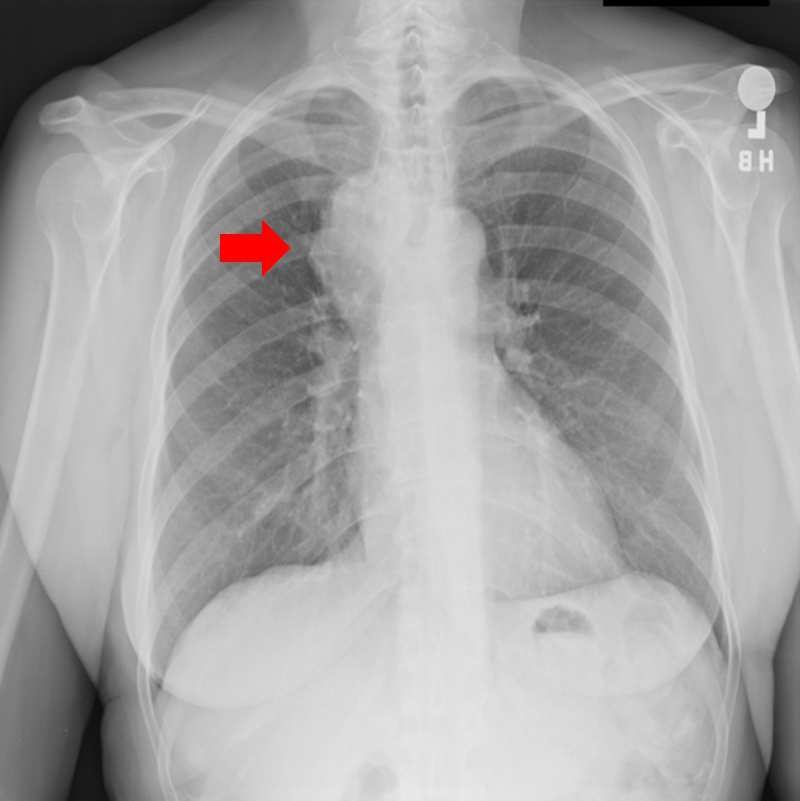}}\\
	\subfigure[Scoliosis]{\includegraphics[width=4cm]{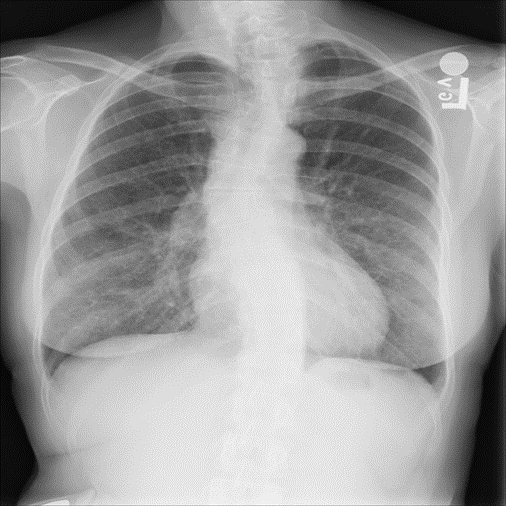}} 
	\subfigure[Multiple bony lesions]{\includegraphics[width=4cm]{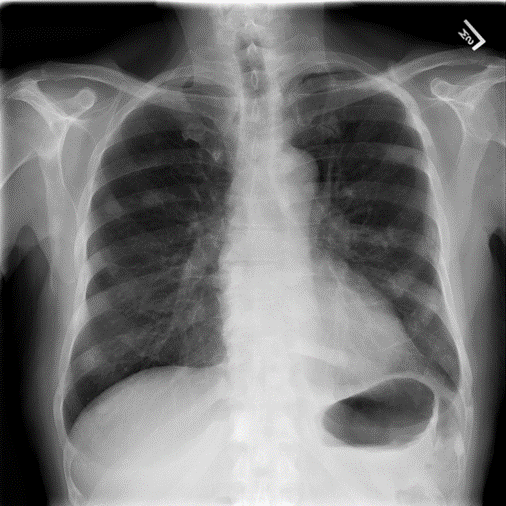}}
	\subfigure[Pneumothorax]{\includegraphics[width=4cm]{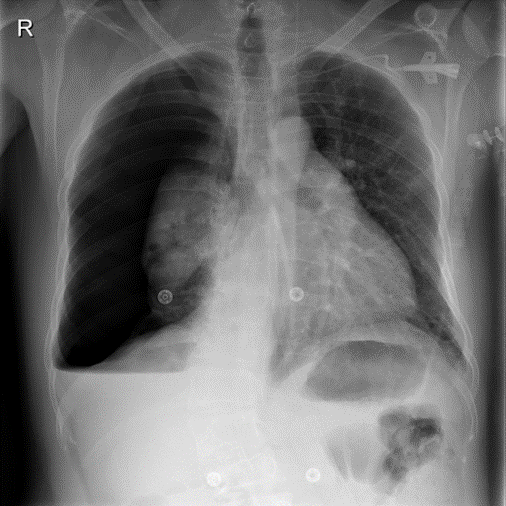}} \\
	\subfigure[Heart~enlargement]{\includegraphics[width=4cm]{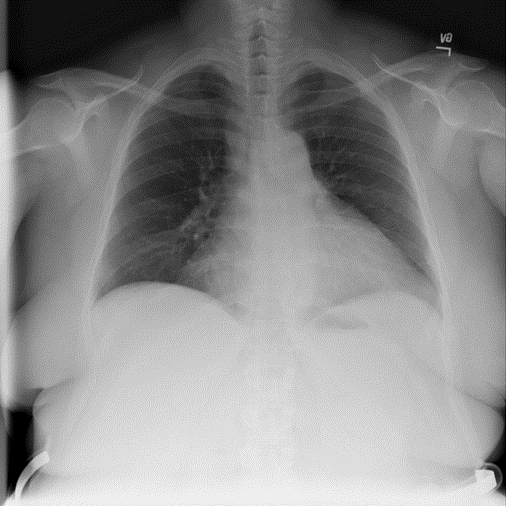}}
	\subfigure[Pneumoperitoneum]{\includegraphics[width=4cm]{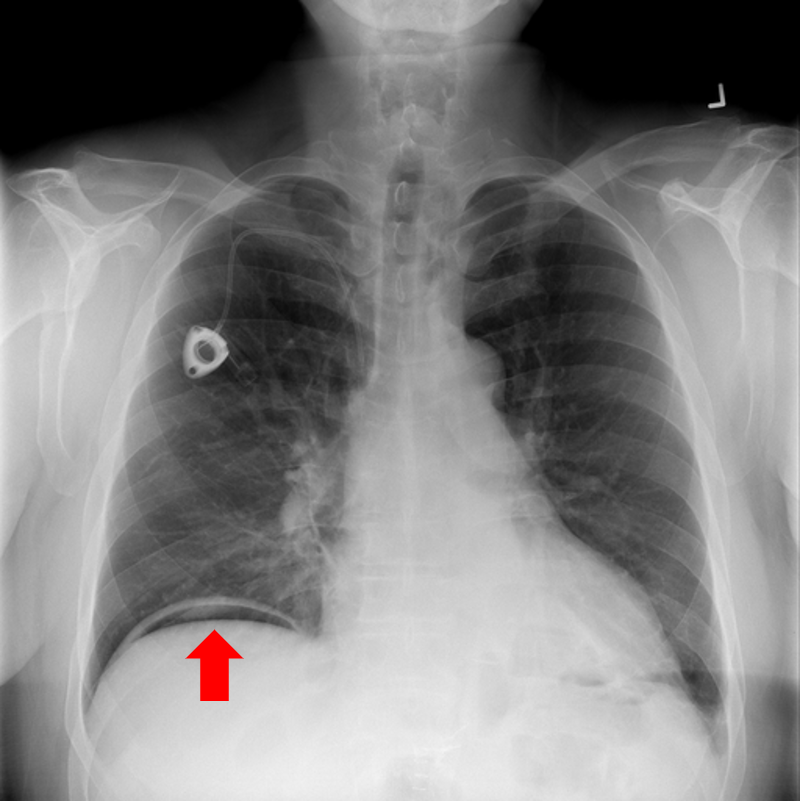}}
	\vspace{-0.2cm}
	\caption{Typical CXRs with a lesion from among the top 1,000 CXRs most likely to have an abnormality according to the logarithm posterior probability out of the 13,863 abnormal CXRs excluding those with a pneumonia-like lung opacity. The lesion is indicated with a red arrow if applicable.}
	\label{fig:others}
\end{figure}
Figure~\ref{fig:others} shows CXRs suggested to have an abnormality different from pneumonia-like lung opacity with the logarithm posterior probability.
These CXRs are the top 1,000 CXRs most likely to have an abnormality out of the 13,863 abnormal CXRs.

\begin{figure}
    \centering
    \includegraphics[width=13.5cm]{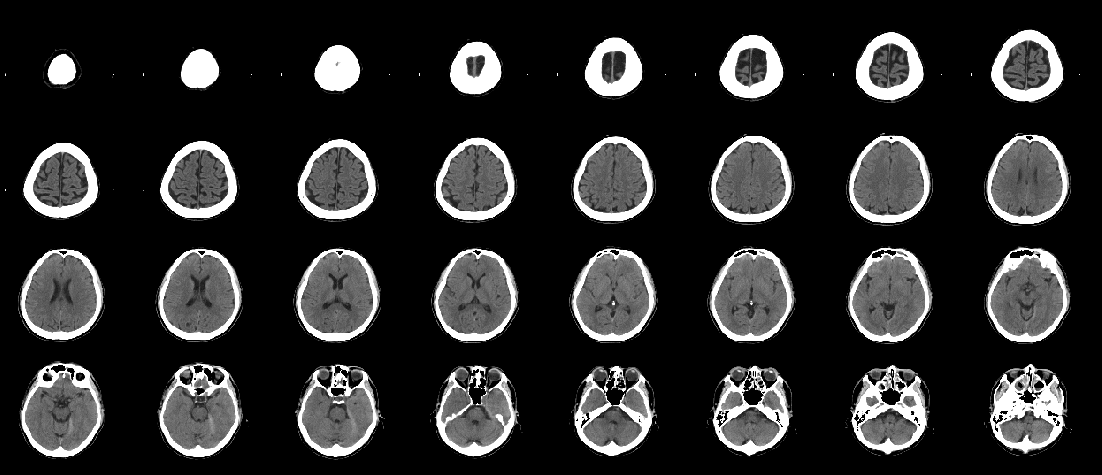}
	\caption{BCT most likely to have an abnormality according to the logarithm posterior probability out of the 33 abnormal BCTs used for testing (subdural hematoma along the left cerebellar tentorium).}
	\label{fig:ctb_detected}
\end{figure}
\begin{figure}
    \centering
    \includegraphics[width=13.5cm]{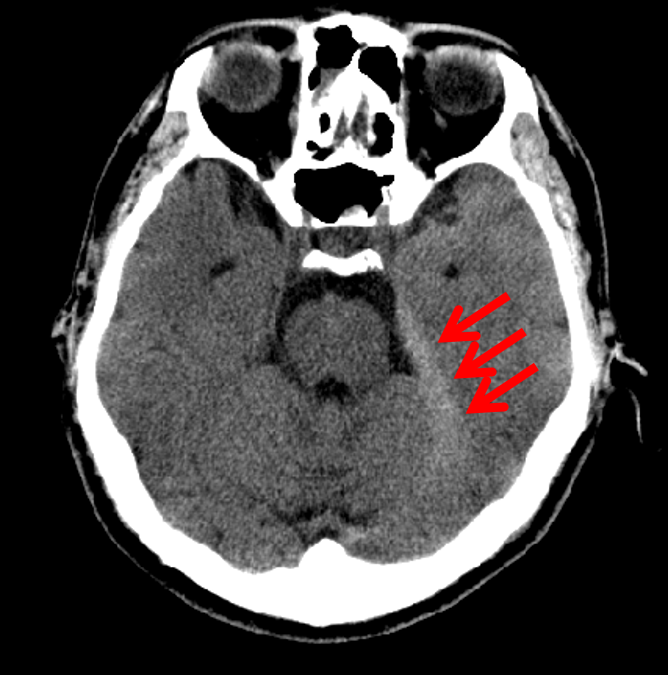}
	\caption{BCT most likely to have an abnormality according to the logarithm posterior probability out of the 33 abnormal BCTs used for testing. This is an enlarged version of a part of Fig.~\ref{fig:ctb_detected}, and the lesion is pointed out with red arrows.}
	\label{fig:ctb_detected_enl}
\end{figure}

\begin{figure}
    \centering
    \includegraphics[width=13.5cm]{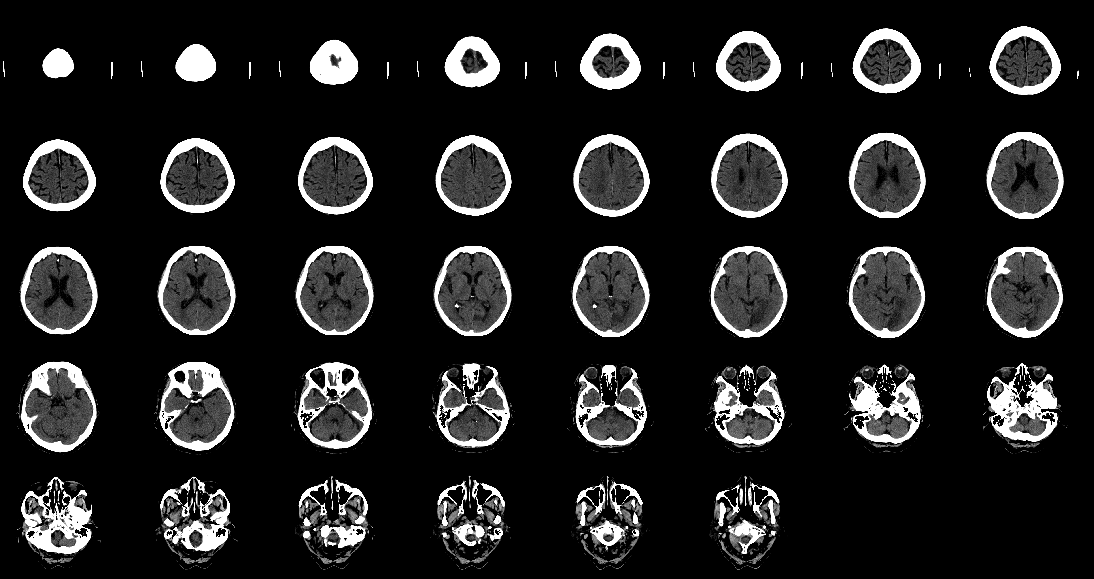}
	\caption{BCT second most likely to have an abnormality according to the logarithm posterior probability out of the 33 abnormal BCTs used for testing (acute infarction in the left posterior cerebral artery territory).}
	\label{fig:ctb_detected2}
\end{figure}
\begin{figure}
    \centering
    \includegraphics[width=13.5cm]{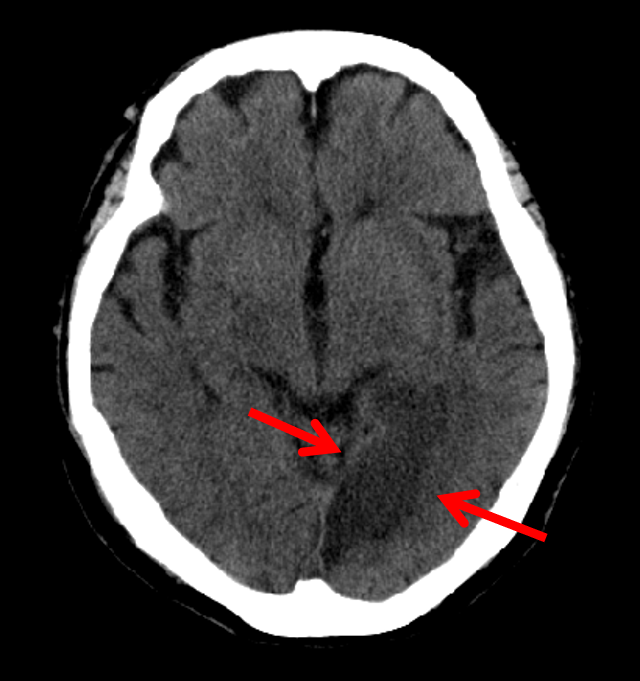}
	\caption{BCT second most likely to have an abnormality according to the logarithm posterior probability out of the 33 abnormal BCTs used for testing. This is an enlarged version of a part of Fig.~\ref{fig:ctb_detected2}, and the lesion is pointed out with red arrows.}
	\label{fig:ctb_detected2_enl}
\end{figure}

Figures~\ref{fig:ctb_detected} and \ref{fig:ctb_detected_enl} show the BCT most likely to have an abnormality according to the logarithm posterior probability out of the 33 abnormal BCTs used for testing.
Figures~\ref{fig:ctb_detected2} and \ref{fig:ctb_detected2_enl} show the BCT second most likely to have an abnormality according to the logarithm posterior probability out of the 33 abnormal BCTs used for testing.
\begin{figure}
    \centering
    \includegraphics[width=13.5cm]{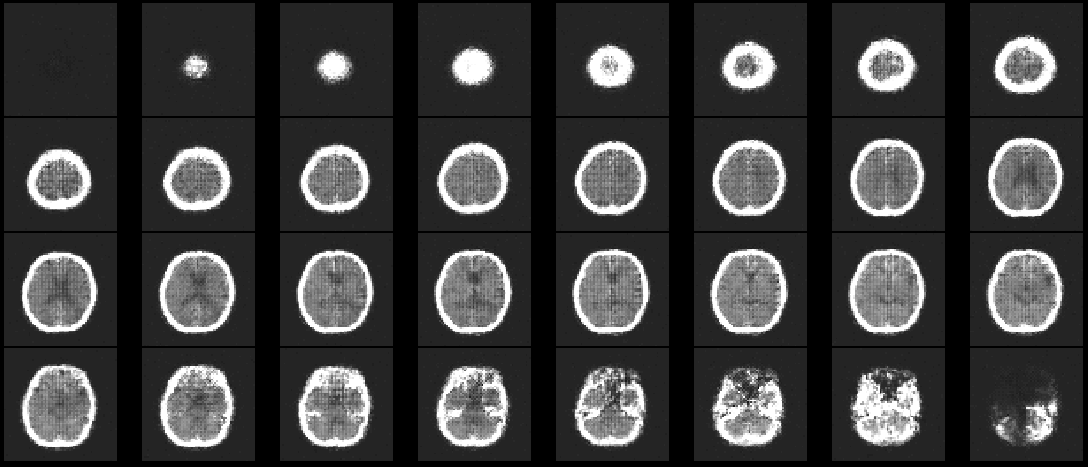} 
    \caption{Fictional image of a BCT generated with the flow-based generative model.}
    \label{fig:imaginary_ctb}
\end{figure}
Finally, a fictional image generated for BCTs is shown in Fig.~\ref{fig:imaginary_ctb}.

\section{Discussion}
\label{sec:discussion}
There has been no previous study on anomaly detection methods for CXRs and/or BCTs in which a flow-based generative model was applied to detect anomalies, to the best knowledge of the authors.
With the present method, it is possible to evaluate the logarithm posterior probability ($\log p \left( \mathrm{normal}|\bm{x}\right)$) of a given CXR or BCT by preparing only two models trained with two datasets.
One dataset is composed of normal medical images, and the other dataset is a mixed set of normal and non-normal medical images.
The preparation of these two datasets is comparatively easy in clinical environments.
To prepare the normal dataset, only medical images that are easily recognized as normal are selected.
To prepare the mixed dataset, a large number of arbitrary CXRs or BCTs obtained in hospitals is sufficient.
Moreover, we do not have to label the data in the mixed dataset.
With these datasets, it is possible to reduce the workload of radiologists and physicians compared with that in supervised settings when they construct a computer-aided diagnosis/detection (CAD) system.


The supervised discriminative model attains a comparatively high AUC, e.g., $\mathrm{AUC} \approx 0.98$ for CXRs \citep{tang2020automated} and $\mathrm{AUC} \approx 0.99$ for BCTs \citep{kuo2019expert}, in exchange for a comparatively high workload for the labeling.
The one-class classifier proposed by Tang et al. \citep{Tang2019, Tang2019b}, in which only normal CXRs are utilized in training, outputted an $\mathrm{AUC}$ of $0.841$ with the Chest X-ray 8 dataset \citep{wang2017chestx}.
Our results for CXRs ($\mathrm{AUC}\approx 0.783$ for overall pathologies, $\mathrm{AUC}\approx 0.868$ for pneumonia-like lung opacity) are lower than that of the discriminative model but are comparable to that of the generative model in previous studies, although a direct comparison of the results obtained with the two different datasets \citep{shih2019augmenting,wang2017chestx} is difficult.

Because our method has no constraints on the logarithm likelihood estimator, it is possible to adopt another flow-based generative model different from Glow.
If one of the most recent flow-based generative models, such as i-ResNet \citep{pmlr-v97-behrmann19a} or residual flows \citep{chen2019residual}, is applied instead of Glow, a higher computational cost is expected.
On the other hand, with such a method, since the learning capability is improved, a higher AUC can also be expected.
The results of adopting different estimators will be included in future works.

In this study, we did not make assumptions on the medical images; hence, we believe that the present method can be easily applied to other medical images, including two-dimensional data obtained by mammography and ultrasound imaging, and three-dimensional data obtained by magnetic resonance imaging and positron emission tomography.
However, note that limitations on the memory space available for GPUs could also limit the applicability of the method.

CAD systems with which three-dimensional medical data are handled require a considerably higher workload for the labeling than those with two-dimensional medical data.
It is expected that the robustness and feasibility, and thus the applicability, of CAD systems are enhanced by using the proposed method because (i) it can potentially handle unknown pathologies and (ii) the workload for the labeling can be greatly reduced.

Our present study has several limitations.
First, the performance (AUC) is inferior to that of the most recent discriminative model with supervised learning \citep{tang2020automated}.
However, there is a possibility that the AUC of the present method can approach that of the supervised discriminative models by greatly increasing the number of medical images used for the training.
Moreover, the preparation of comparatively large-scale datasets for the training can be realized with comparatively low workloads owing to the characteristics of the normal and mixed datasets.
Second, it is currently impossible to indicate the positions of pathological lesions with the present method, and which should be addressed in future works.

\section{Conclusions}
\label{sec:conclusions}
We proposed an anomaly detection method based on flow-based generative models with applicability to both CXRs and BCTs.
With the method, the logarithm posterior probability, i.e., $\log p(\mathrm{normal}|\bm{x}_i)$, was computed using a set of normal CXRs or BCTs and a mixed set of normal and abnormal ones.
For the CXRs, we tested this method using the RSNA Pneumonia Detection Challenge dataset, and the method successfully detected abnormalities with an AUC of 0.783 for overall pathologies and an AUC of 0.868 for pneumonia-like lung opacity.
For the BCTs, we tested the method using images collected from the emergency department of our institution, and the method successfully detected abnormalities with an AUC of 0.866 for overall pathologies, an AUC of 0.904 for infarction, and an AUC of 0.857 for hematoma.
It is concluded that the method can detect abnormal findings in CXRs or BCTs with both an acceptable performance for the testing and a comparatively low workload for the labeling.
The adoption of the method for other medical images is expected in the future.

\section*{Acknowledgements}
Department of Computational Diagnostic Radiology and Preventive Medicine, The University of Tokyo Hospital, is sponsored by HIMEDIC Inc. and Siemens Healthcare K.K.
This work was supported in part by JSPS Grants-in-Aid for Scientific Research KAKENHI Grant Nos. 18K12095 and 18K12096.
This work was also supported by the Joint Usage/Research Center for Interdisciplinary Large-scale Information Infrastructures and High Performance Computing Infrastructure projects in Japan (Project IDs: jh190047-DAH and jh200042-DAH).

\appendix
\section{Results for anomaly detection with the logarithm likelihood}
\begin{figure}
    \centering
    \includegraphics[keepaspectratio,scale=0.35]{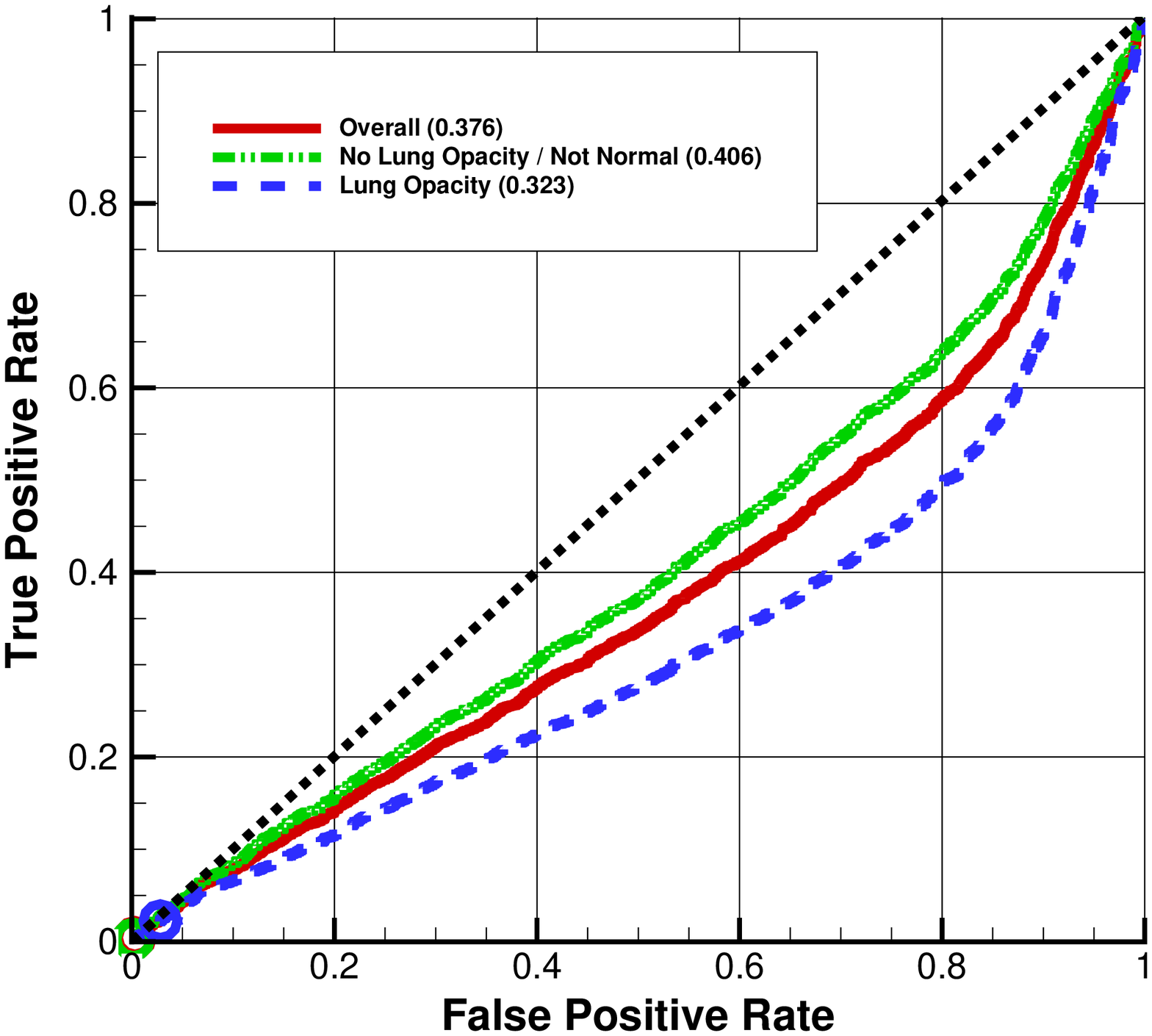}
    \caption{ROC curves obtained with the logarithm likelihood for CXRs. AUCs for the labels are indicated in brackets, and the circles represent cutoff points obtained from Youden's index.}
    \label{fig:lh_only}
\end{figure}

\begin{figure}
    \centering
    \includegraphics[keepaspectratio,scale=0.35]{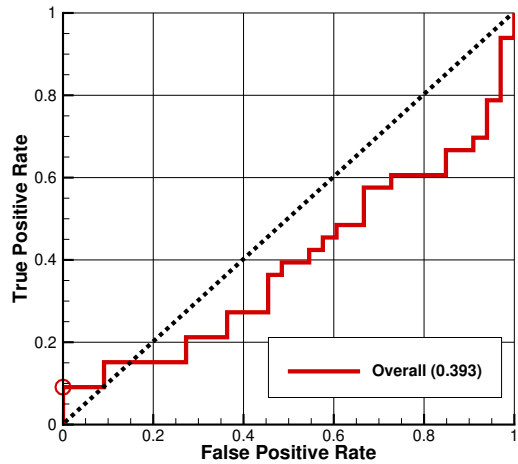}
    \caption{ROC curve obtained with the logarithm likelihood for BCTs. The AUC for the label is indicated in brackets, and the circle represents the cutoff point obtained from Youden's index.}
    \label{fig:lh_only_ctb}
\end{figure}

ROC curves obtained with the logarithm likelihood ($\log p\left(\bm{x}_i | \mathrm{normal}\right)$) with different pathologies are shown in Fig.~\ref{fig:lh_only} for CXRs and in Fig.~\ref{fig:lh_only_ctb} for BCTs.
For each curve, the AUC is less than 0.5; hence, normal cases are likely to be judged as cases with low normality.
These paradoxical results have the same tendency as those in \citep{Ren2019}, where Fashion-MNIST was utilized as the in-distribution dataset (corresponding to normal CXRs or BCTs in this study) and MNIST was utilized as the out-of-distribution dataset (corresponding to non-normal CXRs or BCTs in this study).

\begin{figure}
    \centering
    \includegraphics[width=7cm]{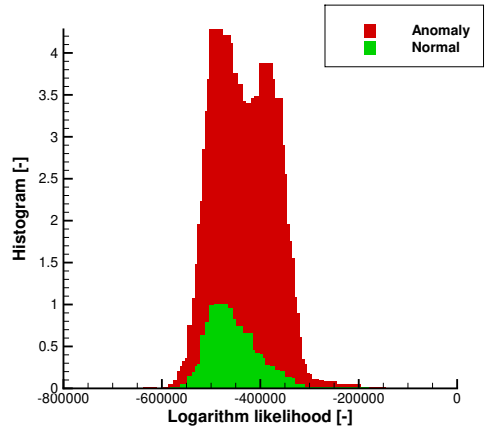}
    \caption{Histogram of the logarithm likelihood for CXRs.}
    \label{fig:cxr_hist_like}
\end{figure}

\begin{figure}
    \centering
    \includegraphics[width=7cm]{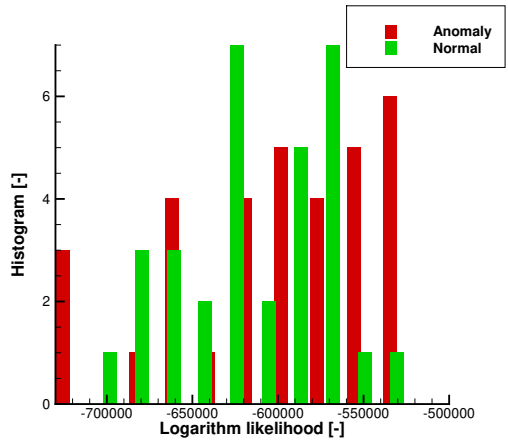}
    \caption{Histogram of the logarithm likelihood for BCTs.}
    \label{fig:bct_hist_like}
\end{figure}

\begin{figure}
	\centering
	\subfigure[]{\includegraphics[width=6.5cm]{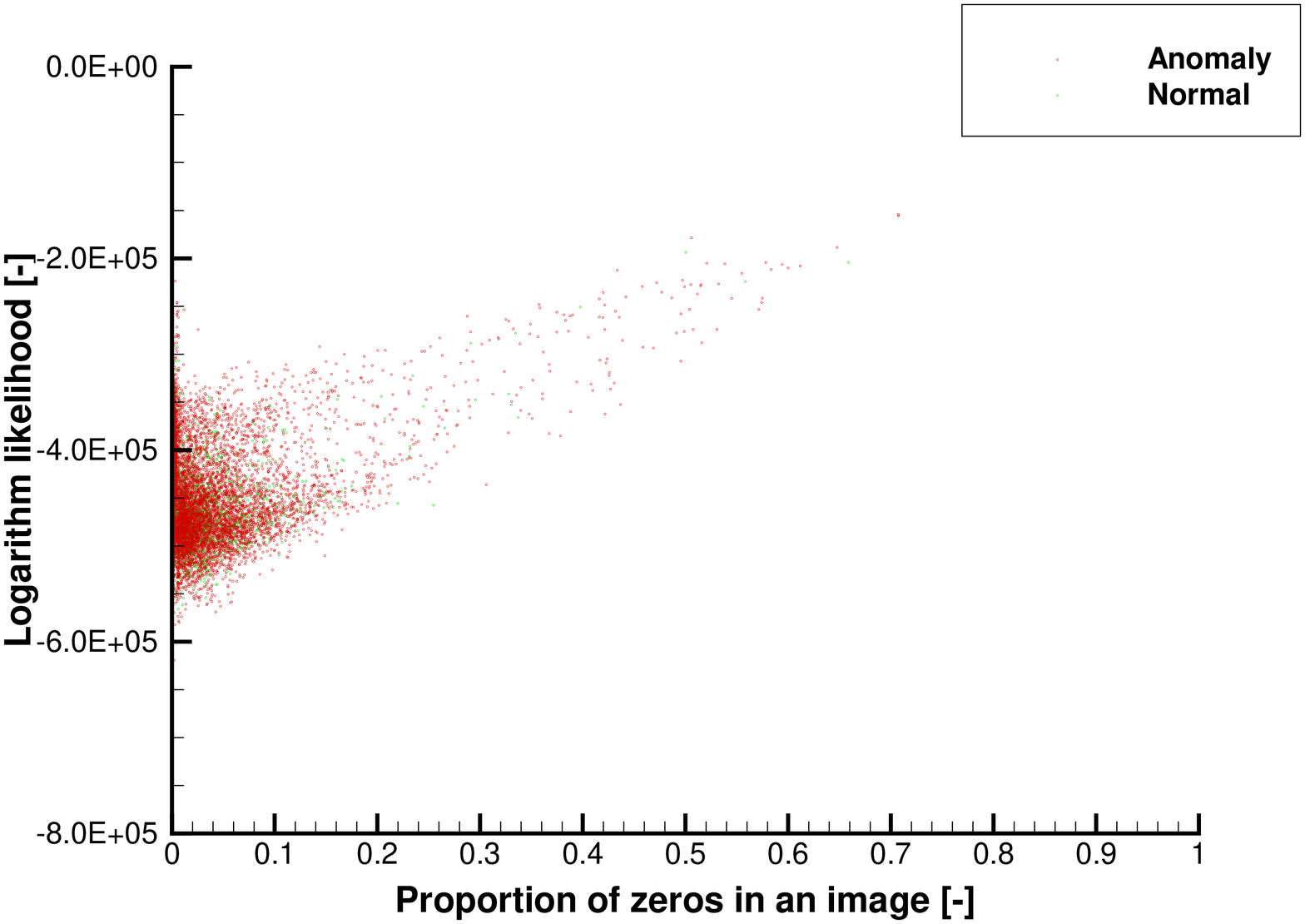}}
	\subfigure[]{\includegraphics[width=6.5cm]{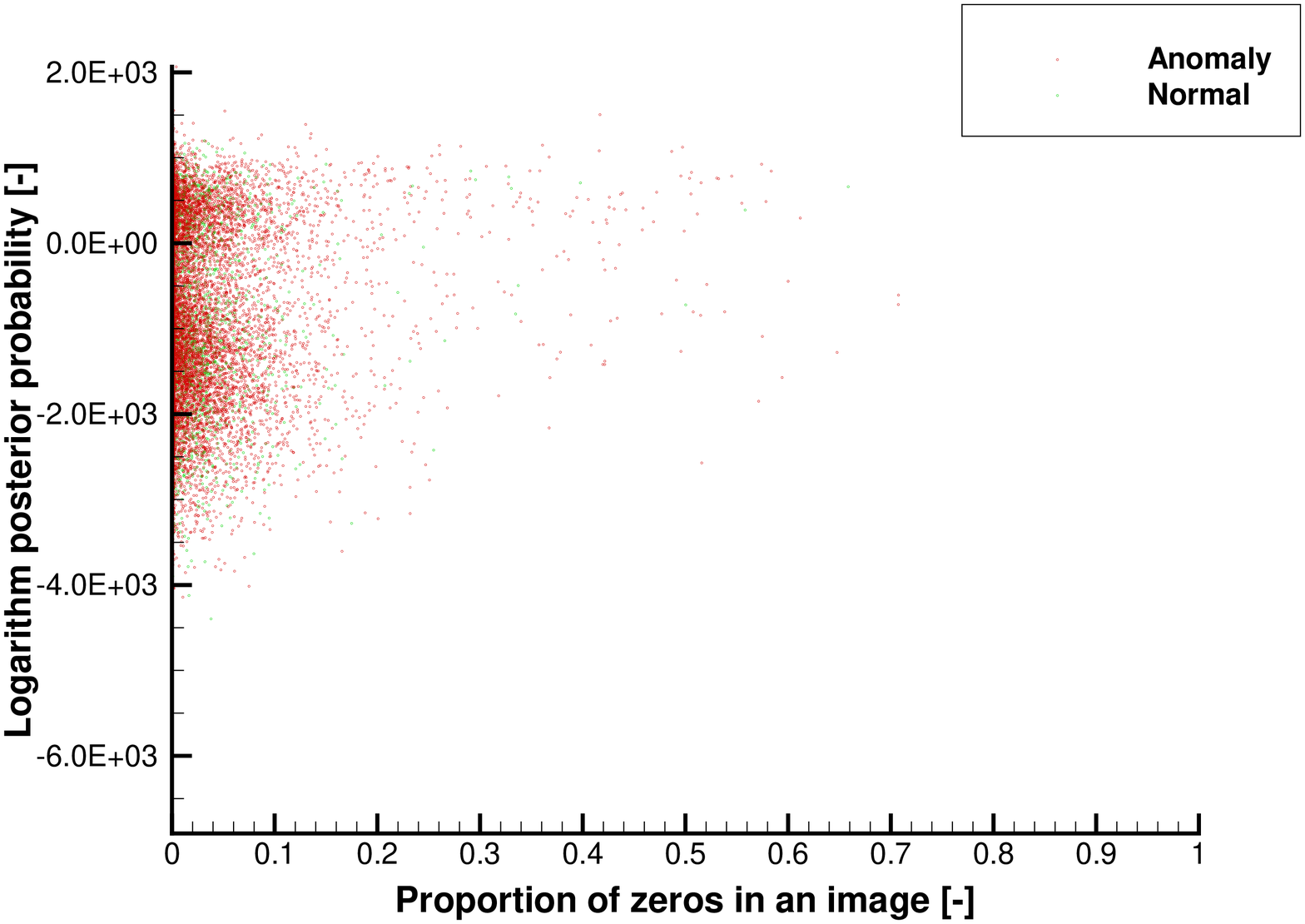}} \\
	\subfigure[]{\includegraphics[width=6.5cm]{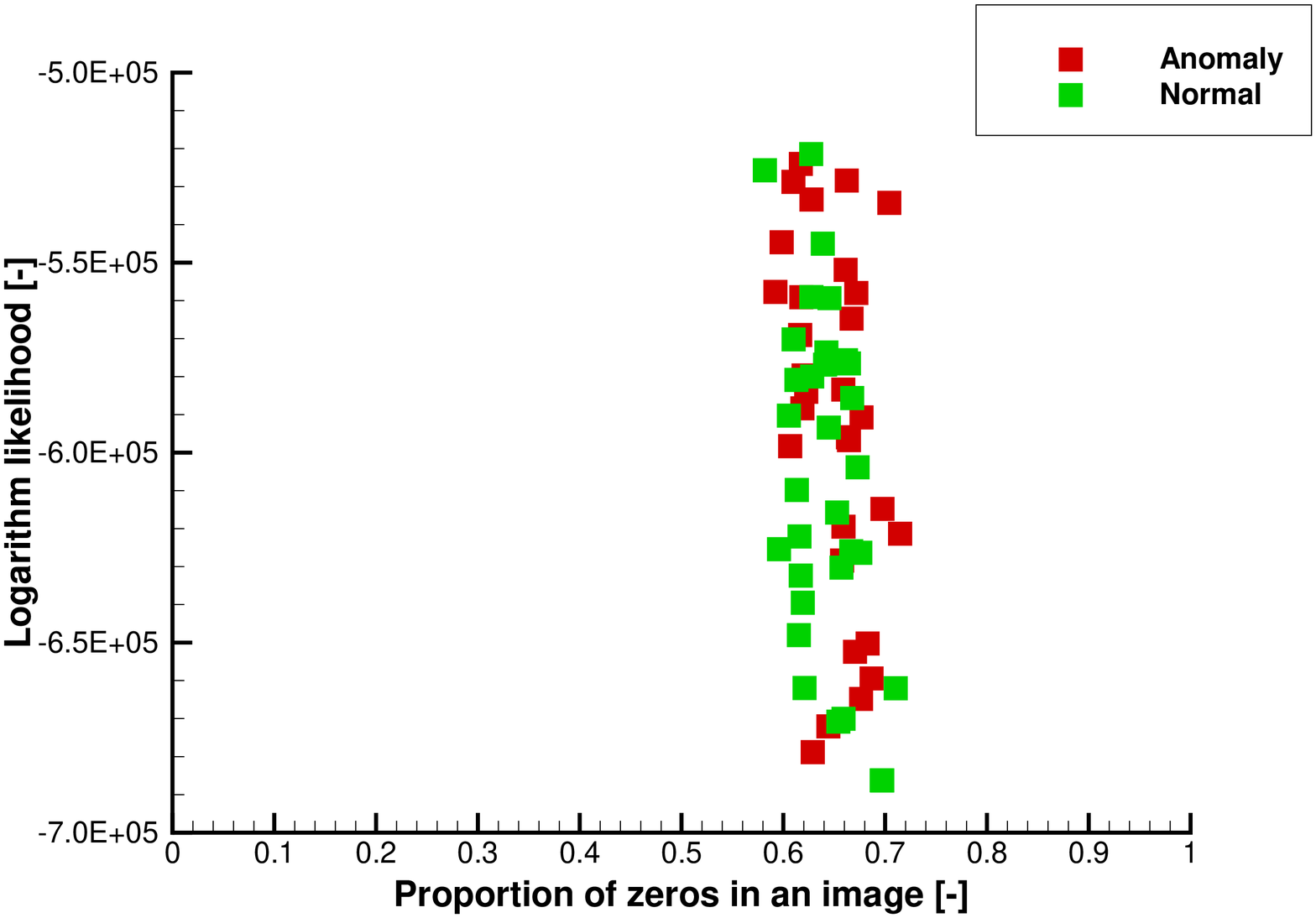}}
	\subfigure[]{\includegraphics[width=5.4cm]{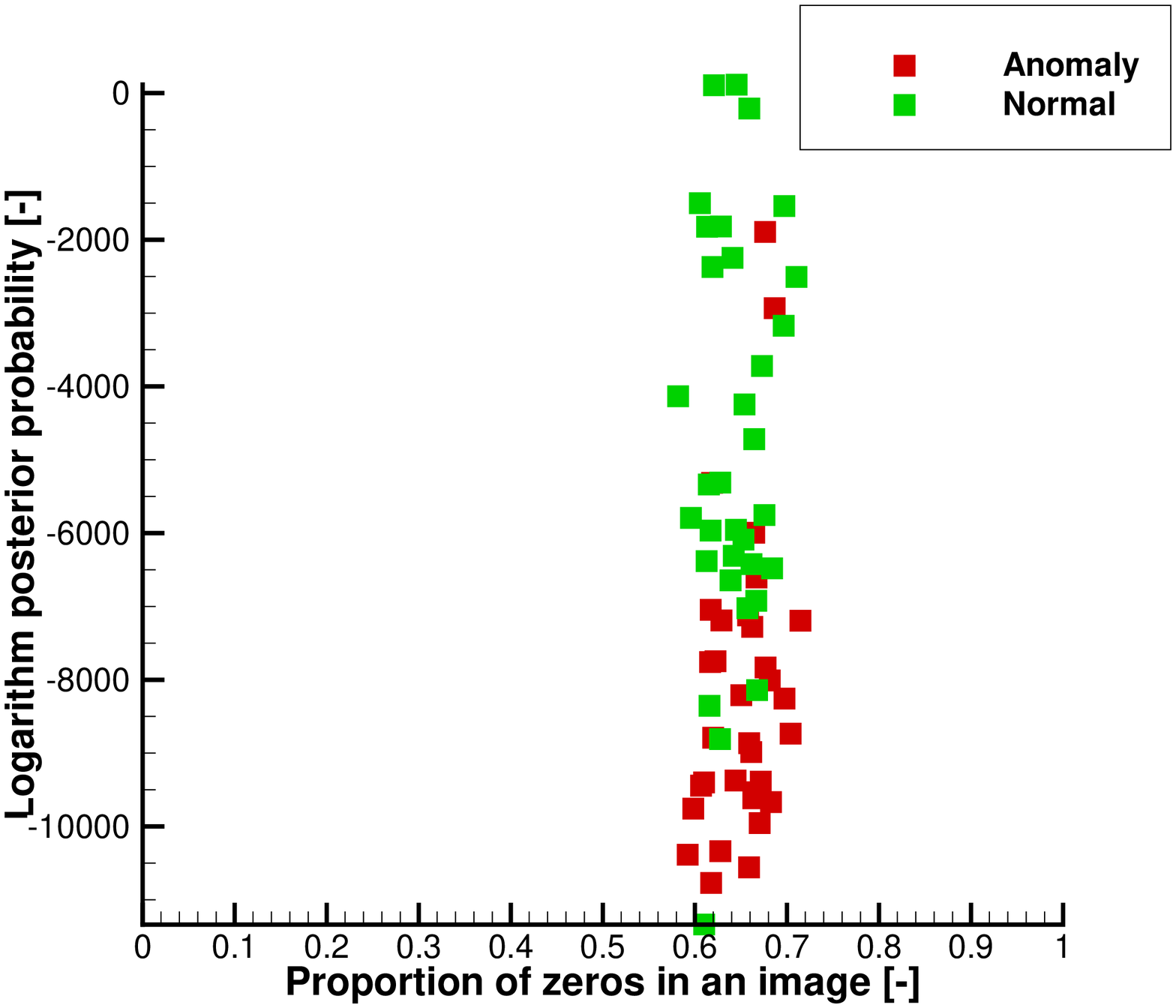}} 
	\vspace{-0.2cm}
	\caption{(a) Relation between the logarithm likelihood and the proportion of zero pixels in a CXR. (b) Relation between the logarithm posterior probability and the proportion of zero pixels in a CXR. (c) Relation between the logarithm likelihood and the proportion of zero pixels in a BCT. (d) Relation between the logarithm posterior probability and the proportion of zero pixels in a BCT.}	
	\label{fig:samp}
\end{figure}

Histograms of the logarithm likelihood are shown in Fig.~\ref{fig:cxr_hist_like} for CXRs and in Fig.~\ref{fig:bct_hist_like} for BCTs.
The relations between the logarithm posterior probability or logarithm likelihood, and the proportion of zero pixels in a CXR or BCT are shown in Fig.~\ref{fig:samp}. 
There is a correlation between the proportion of zero pixels in a CXR and the logarithm likelihood; a CXR with many zero pixels tends to be judged as having higher normality.
This means that background pixels in a CXR have a high impact on the logarithm likelihood, as pointed out in \citep{Ren2019}.
However, there is no apparent correlation observed when the logarithm posterior probability is adopted as the normality metric.

\begin{figure}
    \centering
    \includegraphics[width=14.0cm]{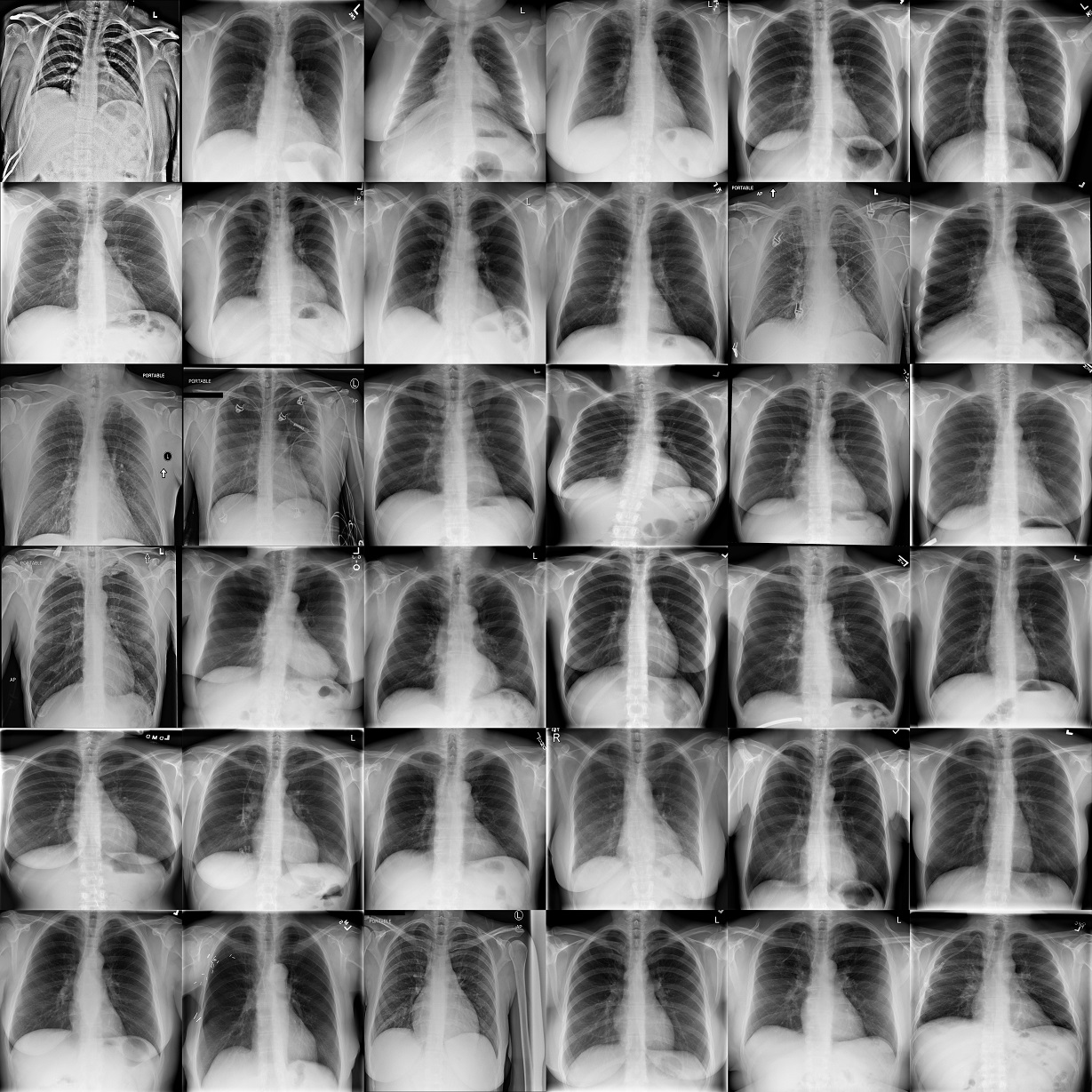}
    \caption{Top 36 CXRs most likely to have an abnormality according to the logarithm likelihood.}
    \label{fig:multi1of4}
\end{figure}

\begin{figure}
    \centering
    \includegraphics[width=14.0cm]{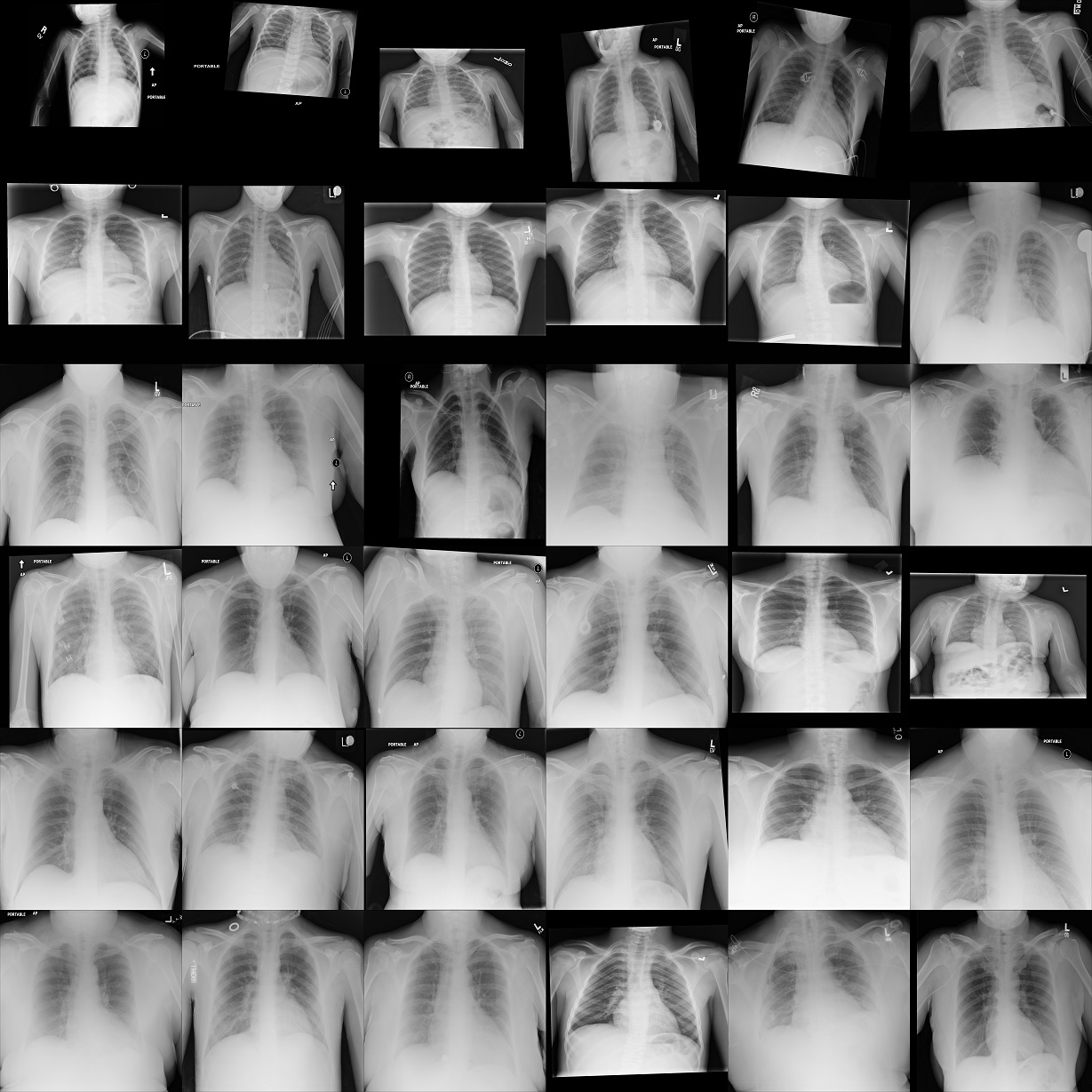}
    \caption{Top 36 CXRs least likely to have an abnormality according to the logarithm likelihood.}
    \label{fig:multi2of4}
\end{figure}

Figures~\ref{fig:multi1of4} and \ref{fig:multi2of4} show the top 36 CXRs most and least likely to have an abnormality from all the test CXRs according to the logarithm likelihood, respectively.
With the logarithm likelihood, a CXR of a child tends to be recognized as a CXR with higher normality, whereas this issue does not arise for the logarithm posterior probability.
The CXR of a child tends to have many zero pixels, and this is considered to be the primary reason for this issue.

\section{Three-dimensional extension of Glow (3D-Glow)}
\subsection{Invertible 1$\times$1$\times$1 convolution}
The invertible 1$\times$1 convolution is mathematically expressed as follows:
\begin{equation}
    x_{\mathrm{out}}\left( h,w,c\right) = \sum_{c'=0}^{C-1} x_{\mathrm{in}} \left( h,w,c' \right) \cdot K\left(c',c\right),
\end{equation}
where $K$ represents a kernel and must be a non-singular square matrix to guarantee invertibility, and
$x_{\mathrm{out}}$ and $x_{\mathrm{in}}$ are the input and output of the convolution procedure, i.e., an image expressed in latent space, respectively.
The indices $h$, $w$, and $c$ run from $0$ to $H-1$, $W-1$, and $C-1$, respectively.
Furthermore, $H$, $W$, and $C$ are the height, width, and channel size of a given image, respectively.
This convolution is computationally realized by inputting a learned kernel whose size is $\left[1,1,C,C\right]$ to \texttt{conv2d} in Tensorflow.

If a dimension in the depth direction is added to the above equation, we have
\begin{equation}
    x_{\mathrm{out}}\left( d,h,w,c\right) = \sum_{c'=0}^{C-1} x_{\mathrm{in}} \left( d,h,w,c' \right) \cdot K\left(c',c\right),
\end{equation}
where the index $d$ runs from $0$ to $D-1$. 
This equation expresses a 1$\times$1$\times$1 convolution.
This convolution is computationally realized by inputting a learned kernel whose size is $\left[1,1,1,C,C\right]$ to \texttt{conv3d} in Tensorflow.

Additionally, the log-determinant of the convolution is computed as
\begin{eqnarray}
    \mathrm{log}\left|\mathrm{det}\left( \frac{\mathrm{d}\texttt{conv3d}\left( {x}_{in}; K \right)}{\mathrm{d}{x}_{in}} \right) \right| = D \cdot H \cdot W \cdot \mathrm{log}\left| \mathrm{det} \left( K \right) \right|.
\end{eqnarray}
The cost of computing $\mathrm{det}(K)$ is $\mathcal{O}(C^3)$, and which is less than the cost of computing $\texttt{conv3d}$ of $\mathcal{O}(D\cdot H\cdot W \cdot C^2)$.

\subsection{Affine coupling layers}
The \texttt{split} routine in Glow \citep{glow_url} is mathematically expressed as follows:
\begin{eqnarray}
    z_1 = z\left(0:H,0:W,0:\frac{C}{2}\right), \\
    z_2 = z\left(0:H,0:W,\frac{C}{2}:C\right), 
\end{eqnarray}
where symbols in the form of ``$a:b$'' mean that the components of a vector are enumerated from $a$ to $b-1$.   
The above equations can be naturally extended to a three-dimensional form as follows:
\begin{eqnarray}
    z_1 = z\left(0:D,0:H,0:W,0:\frac{C}{2}\right), \\
    z_2 = z\left(0:D,0:H,0:W,\frac{C}{2}:C\right).
\end{eqnarray}

\bibliographystyle{elsarticle-harv}
\bibliography{main.bib}

\end{document}